\documentclass[a4paper,12pt]{article}
\pdfoutput=1 

\usepackage{jheppub} 

\usepackage{tabularx,subcaption,fancyhdr,mathtools,physics}
\usepackage[makeroom]{cancel}
\usepackage{bm}
\usepackage{csquotes}
\usepackage[nottoc,notlot,notlof]{tocbibind}
\usepackage{float}
\usepackage{relsize}
\usepackage{hyperref}
\usepackage{color,latexsym, array,multirow,url,verbatim, enumerate,cancel}
\DeclareUnicodeCharacter{2212}{-}
\hypersetup{colorlinks=true,linkcolor=blue,citecolor=magenta,filecolor=magenta,
	urlcolor=cyan}
\def\tcb{\textcolor{blue}}

\def \lspone{\widetilde\chi_1^0}
\def \mlspone{m_{\lspone}}
\def \lsptwo{\widetilde\chi_2^0}
\def \mlsptwo{m_{\lsptwo}}

\def\chonep{\widetilde{\chi}_1^{+}}
\def\chonem{\widetilde{\chi}_1^{-}}
\def\chonepm{\widetilde{\chi}_1^{\pm}}
\def\chonemp{\widetilde{\chi}_1^{\mp}}
\def\mchonepm{m_{\chonepm}}

\def \met{\rm E{\!\!\!/}_T}

\newcommand{\beq}{\begin{equation}}
\newcommand{\eeq}{\end{equation}}
\def\bea{\begin{eqnarray}}
\def\eea{\end{eqnarray}}

\def \lum{\mathcal{L}}

\title{\boldmath Improving sensitivity of trilinear RPV SUSY searches using machine learning at the LHC

}

\author[a]{Arghya Choudhury,}
\author[a]{Arpita Mondal,}
\author[b]{Subhadeep Mondal}
\author[a]{and Subhadeep Sarkar}

\affiliation[a]{Department of Physics, Indian Institute of Technology Patna, Bihar - 801106, India}
\affiliation[b]{Department of Physics, SEAS, Bennett University, Greater Noida, Uttar Pradesh  -201310, India}

\emailAdd{arghya@iitp.ac.in}
\emailAdd{arpita\_1921ph15@iitp.ac.in}
\emailAdd{subhadeep.mondal@bennett.edu.in}
\emailAdd{subhadeep\_1921ph21@iitp.ac.in}

\abstract{In this work, we have explored the sensitivity of multilepton final states in probing the gaugino sector of R-parity violating supersymmetric scenario with specific lepton number violating trilinear couplings ($\lambda_{ijk}$) being non-zero. The gaugino spectrum is such that the charged leptons in the final state can arise from the R-parity violating decays of the lightest supersymmetric particle (LSP) as well as R-parity conserving decays of the next-to-LSP (NLSP). Apart from a detailed cut-based analysis, we have also performed a machine learning-based analysis using boosted decision tree algorithm which provides much better sensitivity. In the scenarios with non-zero $\lambda_{121}$ and/or $\lambda_{122}$ couplings, the LSP pair in the final states decays to $4l~(l = e, \mu) + \met$ final states  with 100\% branching ratio. We have shown that under this circumstance, a final state with $\ge 4l$  
has the highest sensitivity in probing the gaugino masses. We also discuss how the sensitivity can change in the presence of $\tau$ lepton(s) in the final state due to other choices of trilinear couplings. We present our results through the estimation of the discovery and exclusion contours in the gaugino mass plane for both the high luminosity LHC (HL-LHC with $\sqrt{s}=14$ TeV and $\lum = 3000~{\rm fb}^{-1}$) and high energy LHC (HE-LHC with $\sqrt{s}=27$ TeV and $\lum = 3000~{\rm fb}^{-1}$).
For $\lambda_{121}$ and/or $\lambda_{122}$ nonzero scenario, the projected 2$\sigma$ exclusion limit on NLSP masses reaches upto 2.37 TeV and 4 TeV for the HL-LHC and the HE-LHC respectively by using a machine learning based algorithm. 
We observe an enhancement of $\sim$ 380 (190) GeV in the projected 2$\sigma$ 
exclusion limit on the NLSP masses at the 27 (14) TeV LHC. 
Considering the same final state ($N_l \geq 4$) for $\lambda_{133}$ and/or $\lambda_{233}$ non-zero scenario, we find that the corresponding 2$\sigma$ projected limits are $\sim$ 1.97 TeV and $\sim$ 3.25 TeV for the HL-LHC and HE-LHC respectively.  
}

\begin{document} 
\maketitle

\flushbottom

\setlength{\parskip}{1.0em}

\section{Introduction}

Supersymmetry (SUSY) \cite{Martin:1997ns,drees2004theory,baer2006weak} remains one of the most promising candidates for beyond standard model (BSM) physics. This unique extension of the standard model (SM) can address some of the long standing issues in particle physics, such as, the hierarchy problem \cite{SUSSKIND1984181,PhysRevD.14.1667}, the existence of Dark Matter(DM) \cite{Zwicky:1933gu,1937ApJ....86..217Z,Sofue:2000jx,Jungman:1995df}, neutrino oscillation \cite{KamLAND:2013rgu,Borexino:2013zhu,RENO:2018dro,T2K:2018rhz,DayaBay:2018yms,Super-Kamiokande:2019gzr,NOvA:2019cyt}, CP-violation \cite{LHCb:2023tup,BaBar:2015gfp,Belle:2022uod,CDF:2014pzb,Miao:2023cvk}, to name a few. One can also achieve gauge coupling unification \cite{ELLIS1990441,AMALDI1991447,ROSS1992571} within this framework at a higher energy scale. However, no clear evidence of any SUSY particles has yet been obtained from the LHC or other particle physics experiments. This non-observation has led to bounds on these particles \cite{cms_susy, atlas_susy} subjected to the choice of model. The R-parity conserving (RPC) minimal supersymmetric standard model (MSSM) is the most widely studied scenario by the ATLAS and CMS  collaborations \cite{cms_susy, atlas_susy}. Apart from some small pockets of parameter space where the experimental sensitivities are not good enough, the existing LHC data can effectively rule out colored SUSY particle masses upto $\sim$ 2.5 TeV \cite{cms_susy, atlas_susy, ATLAS:2020syg,CMS:2021beq,ATLAS:2021twp}. The bounds are expectedly weaker in the electroweak sector where the lower limits on the SUSY particle masses vary widely \cite{CMS:2021edw,ATLAS:2021yqv,ATLAS:2023lfr,ATLAS:2022hbt,ATLAS:2021moa,CMS:2020bfa,CMS:2019san,CMS:2021cox,CMS:2021ruq} depending on the particle spectrum and available decay modes. The RPC MSSM scenario has the added advantage of a natural DM candidate in the form of the lightest SUSY particle (LSP), but there is no theoretical reason why R-parity cannot be violated.

One of the direct consequences of R-parity violation (RPV) is either lepton number or baryon number violation by one unit. In principle, one can have both lepton number and baryon number violation together, but since that can give rise to proton decay, there are some stringent constraints on the choice of some of the RPV couplings \cite{Dreiner:1997uz}. In presence of R-parity violation, the additional terms one can add in the superpotential are \cite{Dreiner:1997uz,Barbier:2004ez,Banks:1995by}
\begin{equation} \label{eq:rpv_potential}
W_{\cancel{R}_p} = \mu_iH_u.L_i + \frac{1}{2}\lambda_{ijk}L_i.L_je_k^c + \frac{1}{2}\lambda^\prime_{ijk}L_i.Q_jd_k^c + \frac{1}{2}\lambda^{\prime\prime}_{ijk}u_i^cd_j^cd_k^c
\end{equation}
Here L and Q represent left-handed lepton and quark superfields, respectively, and u, d, and e stand for right-handed up quark, down quark, and lepton superfields. $H_u$ denotes up-type Higgs superfield. i, j, k are generation indices and $c$ is charge conjugation. The first three terms in Eq.~\ref{eq:rpv_potential} violate lepton number while the last term is responsible for baryon number violation. Unlike the RPC scenario, the LSP is no more stable and it can decay into SM particles. Therefore, a collider signal of RPC scenario is typically associated with larger missing energy as opposed to RPV signal which has more lepton/jet multiplicity. Depending on the choice of the LSP and non-zero RPV coupling, there can be a plethora of different kinds of final states 
\cite{Dreiner:2023bvs,Barman:2020azo,Mitsou:2015kpa, Bardhan:2016gui,Bhattacherjee:2013tha,Bhattacherjee:2013gr,Dercks:2018eua}.
The experimental collaborations have explored different possibilities to derive limits on the sparticle masses in the context of simplified RPV scenarios \cite{ATLAS:2021fbt,ATLAS:2020wgq,ATLAS:2019fag,ATLAS:2018umm,CMS:2021knz,CMS:2018skt,CMS:2017szl,CMS:2016vuw,CMS:2016zgb,CMS:2013pkf,ATLAS:2015gky,ATLAS:2015rul}.   
One of the major motivations to introduce R-parity violation within SUSY framework is that it can explain neutrino oscillation data \cite{Borzumati:1996hd,Grossman:2003gq,Choudhury:2023lbp,Davidson:2000uc,Mukhopadhyaya:1998xj,Roy:1996bua,Rakshit:1998kd,Allanach:2007qc,Diaz:2014jta,Bose:2014vea,Das:2005mr,Datta:2009dc} which is one of the most robust indications of the existence of BSM physics. In addition to that, RPV SUSY has other advantages, e.g., one can obtain an additional contribution to muon (g-2) \cite{Chakraborty:2015bsk,Altmannshofer:2020axr,Zheng:2021wnu,Chakraborti:2022vds,Maselek:2022cjb,Zheng:2022ssr,Hundi:2011si} or explain flavor anomalies \cite{Bardhan:2021adp,Trifinopoulos:2019lyo,Domingo:2018qfg,Das:2017kfo}. Refer to \cite{Barbier:2004ez} for detailed phenomenological implications of various RPV SUSY scenarios.

As we wait for LHC Run-III to produce more high luminosity data, it is utmost important to assess the impact of the existing data and gauge how much of the new physics parameter space can actually be probed at the highest possible luminosity. 
The gaugino sector of the MSSM is of particular interest because of various phenomenological implications. In RPC context, the gaugino sector is vital in particular from the perspective of 
DM phenomenology \cite{KumarBarman:2020ylm, Barman:2022jdg,He:2023lgi, 
Chakraborti:2015mra, Chakraborti:2014gea, Chakraborti:2017dpu,
Chowdhury:2016qnz,Bhattacharyya:2011se,Choudhury:2012tc,Choudhury:2013jpa} and 
muon (g-2) observation 
\cite{Baer:2021aax,Athron:2021iuf,Endo:2021zal,He:2023lgi,Chakraborti:2021bmv,Choudhury:2017acn,Choudhury:2017fuu, Chakraborti:2015mra, Chakraborti:2014gea, Banerjee:2018eaf,Banerjee:2020zvi,Chakraborti:2021dli,Frank:2021nkq,Ali:2021kxa,Chakraborti:2022vds, Kowalska:2015zja,Chakrabortty:2015ika, Choudhury:2016lku}.
Hence the collider phenomenology of various gaugino production and decay modes have been studied extensively \cite{Choudhury:2016lku,Dutta:2015exw,Chakrabortty:2015ika,Dutta:2017jpe,Barman:2016kgt, Chakraborti:2015mra, Chakraborti:2014gea}. Similar focus on the gaugino sector in the RPV context is somewhat lacking. There can be multiple final states depending on the NLSP-LSP mass gap and available RPV decay modes of the gauginos dictated by the non-zero RPV couplings and their relative strengths compared to the gauge couplings. 
The most stringent constraint on the neutralino-chargino mass plane in RPV context is provided by \cite{ATLAS:2021yyr} through four lepton final state. Looking at the structure of the $\lambda_{ijk}$ coupling, there can be be 9 independent non-zero couplings. 
Depending on the chosen non-zero couplings, one can have varied multiplicities of electrons, muons and taus in the final state. The collider limits are not sensitive to flavors of the leptons as long as only electrons and muons are present in the final state.
If we denote leptons as $l = e, \mu$ then the 9 non-zero couplings lead to 4 different scenarios and we have derived the limits on gaugino masses for all of them. Hence, in this study, we only concentrate on this multi-lepton final state arising from various production channels involving bino-like neutralino LSP and wino-like NLSP. While doing so, we assume that the RPV couplings are large enough such that the LSP decays are prompt. With 
the LHC running almost at full capacity, we need to look not only at the high luminosity option but possible higher center-of-mass (COM) energy options as well. The proposed extension of the COM energy to 27 TeV can be very effective in probing the SUSY scale further. It is necessary to assess what mass range one can effectively probed at this future collider to highlight its importance. 
LHC analyses are steadily moving towards machine learning with the accumulation of more and more data \cite{Alvestad:2021sje,Bhattacherjee:2022gjq,Arganda:2022qzy,Chakraborty:2023hrk,Butter:2022rso}. Algorithms like gradient boosted decision tree (BDT) \cite{Cornell:2021gut,Coadou:2022nsh} can be adopted in collider studies in order to improve on the efficiency of traditional cut-based analyses depending on the suitable choice of kinematical variables. Improved sensitivity towards the new physics signal helps improve projected limits on new physics particle masses. For our analysis, we have adopted the BDT algorithm and used the XGBOOST toolkit \cite{Chen:2016btl}. 

In Sec.~\ref{sec:model} we introduce our model framework and discuss about the possible RPV decay modes of the bino LSP. We also discuss the various possible final state given the decay modes. In Sec.~\ref{sec:collider_analysis} we briefly mention about how the events are reconstructed. In Sec.~\ref{sec:14_cut} we define the signal regions for cut-based analysis for the HL-LHC. In this section, we also show the new projected exclusion limits on chargino and neutralino masses derived through our analysis. After that we repeat the same final state analysis in the Sec.~\ref{sec:14_ml} by using machine learning (ML) algorithm and compare the results with that of cut based analysis. In Sec.~\ref{sec:27_cut} we proceed with our analysis with 27 TeV COM energy in order to find the reach in chargino neutralino mass plane. We again do the ML-based analysis for 27 TeV COM energy and compare the results in Sec.~\ref{sec:27_ml}. Finally, we conclude in Sec.~\ref{sec:conclusion}.

\section{Model Framework}
\label{sec:model}

Among the different electroweakino productions, the wino production cross sections are the most significant one. In this analysis, we consider a simplified RPV SUSY scenario with light winolike $\chonepm$  and $\lsptwo$, which are mass degenerate. The LSP ($\lspone$) is assumed to be pure bino like and due to $\lambda_{ijk}L_iL_j\bar E_k$ coupling it decays as $\lspone \rightarrow  l_k^{\prime\pm}l_{i/j}^{\prime\mp}\nu_{j/i}$ via virtual sneutrino/sleptons, where $l^\prime = e, \mu,$ and $ \tau$. 
We present all the possible decay modes of the LSP $\lspone$ via $\lambda_{ijk} (i,j,k =1,2,3)$ couplings in Table~\ref{tab:br1} for different allowed values of $i,j, $ and $k$\footnote{Due to the gauge invariance of Superpotential the couplings are antisymmetric in first two indices, i.e., $\lambda_{ijk}$ = - $\lambda_{jik}$}. 
For a single non-vanishing coupling, $\lspone$ decays to leptonic ( $l =e/\mu$) final states with 100\% branching ratios for $\lambda_{121}$ and $\lambda_{122}$. On the other hand, $\lambda_{133}$ and $\lambda_{233}$ couplings allow the LSP to decay into tau enriched final states with $1\tau$ and $2\tau$ final states with 50\% branching ratios each (see Table~\ref{tab:br1}). 

\begin{table}[h]
\centering
\begin{tabular}{||c|c|c|c||} 
    \hline \hline
 	 & k = 1 & k = 2 & k = 3 \\
    \hline \hline
 	 ij = 12 & $ee\nu_{\mu}$, $e\mu\nu_{e}$ & $\mu e\nu_{\mu}$, $\mu \mu \nu_e$ & $\tau e\nu_{\mu}$, $\tau \mu \nu_e$ \\
    \hline \hline
 	 ij = 13 & $ee\nu_{\tau}$, $e\tau \nu_e$ & $\mu e \nu_{\tau}$, $\mu \tau \nu_e$ & $\tau e \nu_{\tau}$, $\tau \tau \nu_e$ \\
    \hline \hline
 	 ij = 23 & $e \mu \nu_{\tau}$, $e \tau \nu_{\mu}$ & $\mu \mu \nu_{\tau}$, $\mu \tau \nu_{\mu}$ & $\tau \mu \nu_{\tau}$, $\tau \tau \nu_{\mu}$ \\
    \hline \hline
\end{tabular}
\caption{All possible decay modes of LSP $\lspone$ (with 50\% branching ratios each) corresponding to nine different coupling choices of $\lambda_{ijk}$ assuming only one coupling is non-zero.}
\label{tab:br1}
\end{table}

In the context of RPC SUSY searches, the most studied analysis is $\chonepm \lsptwo$ pair production with $3l + \met$ final states. But for RPV scenarios with LLE operators, signal efficiencies will depend on leptons coming from LSPs, rather than NLSPs ($\chonepm, \lsptwo$). In this analysis, we consider  $\chonepm \lsptwo$ pair production along with $\chonepm \chonepm$ pair production \footnote{We keep the masses of the other SUSY particles likes squarks, gluino, sleptons, heavy Higgses and heavier electroweakinos fixed at beyond 5 TeV.}. Production cross section for the first process is roughly double of the later mode and in both the cases NLSPs dominantly produce via RPC couplings. It may be noted that  $\chonepm \lspone$, $\lspone \lsptwo$, $\lspone \lspone $, $\lsptwo \lsptwo$ production rates are almost vanishing for bino $\lspone$ and winolike $\chonepm$  and $\lsptwo$ \footnote{For example, 
$\sigma(pp \to \chonepm \lsptwo$) = 1.83 fb, $\sigma(pp \to  \chonepm\tilde\chi_1^{\mp}$) = 0.84 fb, $\sigma(pp \to \chonepm \lspone$) or $\sigma(pp \to \lsptwo \lsptwo) \sim 10^{-8}$ fb at 14 TeV LHC with electroweakino mass around 1 TeV.}. In our case $\chonepm$ decays into $W^\pm \lspone$ and $\lsptwo$ can decay into  $Z \lspone$ and/or $h \lspone$. For the sake of simplicity we assume that Br($\lsptwo \to Z \lspone$) = Br($\lsptwo \to h \lspone$) = 50\%. From $\chonepm \lsptwo$ and $\chonepm \chonepm$ pair productions final state always consists of a LSP pair and the lepton multiplicity in the final states depends on the choices of non-zero $\lambda_{ijk}$. In Table~\ref{tab:br2}, we summarize the charged lepton configuration coming from a LSP pair for only one single non-zero choice of $\lambda_{ijk}$.

\begin{table}[h]
\begin{center}
\begin{tabular}{|c|c|c|} 
    \hline \hline
 	Non-zero  & Charged lepton configuration & Remarks  \\
 	couplings & (Branching Ratios) & ($l~=~e,\mu$ only) \\
    \hline \hline
 	$\lambda_{121}$ & ~~~ $4e$(25\%) ~~~ ~~~ $3e1\mu$(50\%) ~~~ ~~~ $2e2\mu$(25\%) &  $4l$ (100\%) \\
 	$\lambda_{122}$ & ~~~ $4\mu$(25\%) ~~~ ~~~ $3\mu1e$(50\%) ~~~ ~~~ $2e2\mu$(25\%) & \tcb {\textbf{Scenario-I}}\\
    \hline
    \hline
     $\lambda_{131}$ &   ~~~  $4e$(25\%) ~~~ ~~~ $3e1\tau$(50\%) ~~~ ~~~ $2e2\tau$(25\%) & $4l$(25\%) \\
     $\lambda_{232}$ &   ~~~  $4\mu$(25\%) ~~~ ~~~ $3\mu 1\tau$(50\%) ~~~ ~~~ $2\mu 2\tau$(25\%) & $3l1\tau$(50\%) \\
     $\lambda_{132}$ &   ~~~ $2\mu 2e$(25\%) ~~$1e2\mu 1\tau$(50\%) ~~~~~~ $2\mu 2\tau$(25\%) & $2l2\tau$(25\%) \\
     $\lambda_{231}$ & ~~~ $2e2\mu$(25\%)~~~ $2e1\mu 1\tau$(50\%) ~~~~~~ $2e2\tau$(25\%) & 
     \tcb {\textbf{Scenario-II}}  \\
    \hline
    \hline
        $\lambda_{123}$ &  ~ $2e2\tau$(25\%)~~~~~$1e1\mu 2\tau$(50\%) ~~~~~ ~$2\mu 2\tau$(25\%) & $2l2\tau$(100\%) \\
    				&					& 	\tcb {\textbf{Scenario-III}}		\\
    				\hline
    				\hline
    $\lambda_{133}$ &  $2e2\tau$(25\%) ~~~ $1e3\tau$(50\%) ~~~~~~~~~~~~ $4\tau$(25\%) & $2l2\tau$(25\%) \\
    $\lambda_{233}$ & $2\mu 2\tau$(25\%) ~~~ $1\mu 3\tau$(50\%) ~~~~~~~~~~~ $4\tau$(25\%) & $1l3\tau$(50\%)\\
         &  & $4\tau$(25\%) \\
             				&					& 	\tcb {\textbf{Scenario-IV}}		\\
\hline
    \hline
\end{tabular}
\caption{Charged lepton configuration arises from a LSP pair in the  RPV LLE scenarios for  a single non zero $\lambda_{ijk}$ coupling. Four scenarios ({\textbf{Scenario-I to Scenario-IV}}) are defined according to the leptonic branching ratios for further analysis.}
\label{tab:br2}
\end{center}
\end{table}

It is evident from Table~\ref{tab:br2} that if either $\lambda_{121}$ or $\lambda_{122}\ne 0 $ then the LSP pair gives $4l~(l = e, \mu) + \met$ final states  with 100\% branching ratios (defined as {\textbf{Scenario-I}}). It was shown in Ref~\cite{ATLAS:2014pjz} that the mass limits are almost similar for $\lambda_{121} \ne 0$ and 	$\lambda_{122} \ne 0$ scenarios i.e., the distinction between electron and muon is not sensitive/essential in the $4l~(l = e, \mu) + \met$ final states for obtaining the exclusion limits. It may be noted that when both $\lambda_{121}$ and $\lambda_{122}$ are non-zero, LSP decays via $e^{\pm}e^{\mp}\nu_{\mu}$(25\%), $e^{\pm}\mu^{\mp}\nu_{\mu}$(50\%) and $\mu^{\pm}\mu^{\mp}\nu_e$(25\%), but the LSP pair contributes to $4l$ states similar to {\textbf{Scenario-I}}. Other extreme case ({\textbf{Scenario-IV}}) is obtained for $\lambda_{133} \ne 0$ or $\lambda_{233} \ne 0$, where tau enriched  final states $2l2\tau$(25\%), $1l3\tau$(50\%) and $4\tau$(25\%) emerge from the LSP pair. If both $\lambda_{133}$ and $\lambda_{233}$ are non-zero then we also get the same final states. For other options of $\lambda_{ijk}$, the branching fractions of leptonic final states lie 
between these two extreme scenarios and are summarized in Table~\ref{tab:br2}. If both $\lambda_{121}$ and $\lambda_{133}~\neq~0$, then $\lspone$ decay via $e^{\pm}e^{\mp}\nu_{\mu}$, $e^{\pm}\mu^{\mp}\nu_{\mu}$, $\tau^{\pm} e^{\mp}\nu_{\tau}$ and $\tau^{\pm} \tau^{\mp} \nu_e$ with each 25\% branching fraction and eventually the LSP pair leads to the final states: $4l$(25\%), $3l1\tau$(25\%), $2l2\tau$(31.25\%), $1l3\tau$(12.5\%) and $4\tau$(6.25\%). Thus the discovery reach/exclusion limits for scenarios with a combination of  two or more non-zero $\lambda_{ijk}$ will be achieved between {\textbf{Scenario-I}} and {\textbf{Scenario-IV}}. 


\section{Collider Analysis}
\label{sec:collider_analysis}
 
\begin{figure}[h]
\centering
\includegraphics[width=0.45\linewidth]{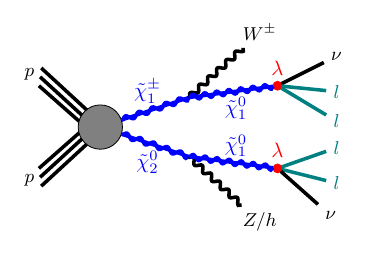}
\includegraphics[width=0.45\linewidth]{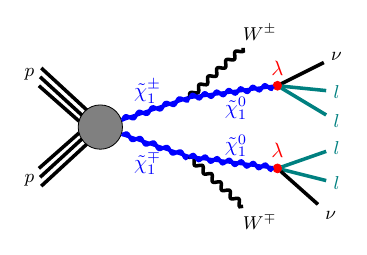}
\caption{Diagrams of wino like NLSP pair $\chonepm \lsptwo$ and $\chonepm \chonemp $ 
productions via RPC electroweak process and the consequent decay of the LSP ($\lspone$) via RPV $LL\bar E$ couplings.}
\label{fig:rpv_decay}
\end{figure}

For our analysis, we consider two production channels $pp\to \chonepm \lsptwo$ and $pp\to \chonep \chonem$. the Feynman diagram for 
these production channels and subsequent decays considered in this analysis are depicted in Fig.~\ref{fig:rpv_decay}. As mentioned 
earlier, in the simplified model the production and decays of NLSP occur via conventional RPC mode and LSP decays promptly via RPV $LL\bar E$ couplings.
Our focus will be to obtain the sensitivity  reach in the mass plane via  detailed collider analysis in the {\textbf{Scenario-I}} 
\footnote{We also discuss about the exclusion reach of our signals in other scenarios mentioned in Table~\ref{tab:br2}.}. 
The leptons in the final state arise from the decay of the LSP as well as from the W/Z boson decay, which originates from the NLSP (illustrated in Fig.~\ref{fig:rpv_decay}). Hence we only consider the final states consisting of atleast four leptons ($N_l \ge 4, \text{where}~ l =e, \mu$).   
 
For the $N_l\geq 4$ channel, the most dominant source of backgrounds are $ZZ+jets$, $WWZ+jets$ and $t\bar{t}Z+jets$. Additional contributions arise from $WWZ+jets$, $ZZZ+jets$, Higgs ($h$) production via gluon gluon fusion (ggF). We have also generated/considered the processes like $hjj$, $Wh+jets$ and $Zh+jets$. All the SM background events have been generated using \texttt{MadGraph5-aMC@NLO} \cite{Alwall:2014hca} at the leading order (LO) parton level. The cross sections of different SM background processes used in this analysis and the generation level cuts have been tabulated/listed in Appendix A (see Table~\ref{tab:cs_bkg_14} and Table~\ref{tab:cs_bkg_27} for 14 TeV and 27 TeV COM energies respectively). 
These events have been generated by matching up to 2 jets (for $ZZ$, 3 jets matched sample is used). The SUSY signal events have been generated using \texttt{Pythia-6.4.28} \cite{Sjostrand:2006za}. The next-to-leading order + next-to-leading logarithmic (NLO+NLL) order cross-sections have been computed for the signal events ($\chonep \chonem$ and  $\chonepm \lsptwo$) by using \texttt{Resummino-3.1.1} \cite{Fuks:2013vua}. The showering and hadronization for background events have been done through  \texttt{Pythia-8} and for the signal events we have used \texttt{Pythia-6.4.28}. Then all these events passed through a fast detector simulation in the 
\texttt{DELPHES 3} platform \cite{deFavereau:2013fsa} (version-3.5.0). Using anti-$k_t$ algorithm \cite{Cacciari:2008gp} algorithm within \texttt{Fastjet} \cite{Cacciari:2011ma}  
framework, jets have been reconstructed with jet radius parameter $R = 0.4$ along with transverse momentum $p_T > 20$ and pseudorapidity range $|\eta| < 2.8$. Following the ATLAS analysis \cite{ATLAS:2021yyr}, the b-tagging efficiency has been chosen to be 85$\%$ and the light jet  mistagging efficiency (as b-jet)  25$\%$. For the identification of $b$-jets,  the pseudorapidity range $|\eta| < 2.5$ has been considered. 

For the reconstruction of leptons (electron, muon), we have followed the isolation, overlap removal procedures etc. according to the ATLAS analysis as mentioned in Sec.~5 of  \cite{ATLAS:2021yyr}. The final state electrons (muons) are required to have $p_T > $ 7 (5) GeV,  $|\eta|<$ 2.47 (2.7), and must satisfy both the track isolation and calorimeter isolation criteria. We have considered the \texttt{Loose} isolation criteria \cite{ATLAS:2019qmc,ATLAS:2020auj} for both the leptons where the conditions on the  scalar sum of $p_T$ of the surrounding particles are ${\sum p_T^{varcone20}}/{p_T^{e(\mu)}} < 0.15~(0.15)$ and 
 ${\sum E_T^{cone20}}/{p_T^{e(\mu)}} < 0.20~(0.30)$ for tracker and calorimeter isolation respectively for selected electrons (muons)\footnote{ Surrounding objects  with $p_T \ge 1.0$ GeV are chosen within  a cone radius of $\Delta R =$  0.2 of leptons for calorimeter isolation. For track isolation a variable cone of min[${10~GeV}/{p_T^e}, 0.2$] and min[${10~GeV}/\{{p_T^{\mu}\}}, 0.3]$ are considered for electron and muon respectively. $\Delta R$ is calculated from  the differences of pseudorapidity and azimuthal angle as $\Delta R = \sqrt{(\Delta \eta)^2 + (\Delta \phi)^2}$.}.
Furthermore to suppress the decays of low mass particles, both leptons are discarded if they form an opposite sign (OS) or a same flavour opposite sign (SFOS) pair with the invariant mass of the pair being $M_{OS}<$ 4 GeV and 8.4 $< M_{SFOS} < 10.4$ GeV respectively. In the next subsections (Sec.~\ref{sec:14_cut} and Sec.~\ref{sec:14_ml} we will first present a detailed cut based collider study for the HL-LHC and then we will study the improvement on the exclusion reach by using machine learning-based methods. In a similar manner, the future prospects for the HE-LHC will be presented in Sec.~\ref{sec:27_cut} and Sec.~\ref{sec:27_ml}.


\subsection{Prospect at the HL-LHC using cut-based analysis}
\label{sec:14_cut}   

In this section we present the search prospect of wino pair production at the High Luminosity LHC (HL-LHC) with $N_l\geq 4$ channel at center-of-mass energy, $\sqrt{s} =$ 14 TeV and luminosity, $\mathcal{L} =$ 3000 $fb^{-1}$ via traditional cut-and-count analysis. Using the Run-II LHC data, the ATLAS collaboration has already excluded wino mass around 1.5 TeV \cite{ATLAS:2021yyr} for relatively large $\mlspone$ and for the cut based analysis we closely follow this analysis. 
We carry our analysis for two signal regions: \texttt{SR-A} and \texttt{SR-B} \footnote{The signal regions are differentiated through different $m_{eff}=\sum_i p_T^{l_i} + \sum_i p_T^{j_i} + \met$ cut as discussed later.} which 
are optimized for smaller and larger masses of $\chonepm$/$\lsptwo$ respectively. For these signal regions, we estimated the signal yields by varying the 
$ \mchonepm(=\mlsptwo)$ in the range 1-3 TeV and $\mlspone$ in the range 
50 GeV to ($\mlsptwo-10$)  GeV\footnote{To assure the prompt decay of $\lspone$.} 
with a step size of 10 GeV. 
We have chosen three signal benchmark points  to showcase our results - 
\texttt{BP1}: $\mchonepm = 1600$ GeV, $\mlspone = 250$ GeV, 
\texttt{BP2}: $\mchonepm = 1800$ GeV, $\mlspone = 800$ GeV, 
\texttt{BP3}: $\mchonepm = 1950$ GeV, $\mlspone = 1850$ GeV. 
These benchmark points are selected on the basis of mass difference between $\mchonepm$ and $\mlspone$  i.e.,  large,  intermediate and small mass differences. 
The details of the background cross-section along with the yield after the generation level cut are summarised in Table~\ref{tab:cs_bkg_14} and the NLO+NLL level cross-sections for the benchmark points have been tabulated in Table~\ref{tab:cs_sig}. 

\begin{figure}[!htb]
\begin{center}
    \includegraphics[width=0.8\textwidth]{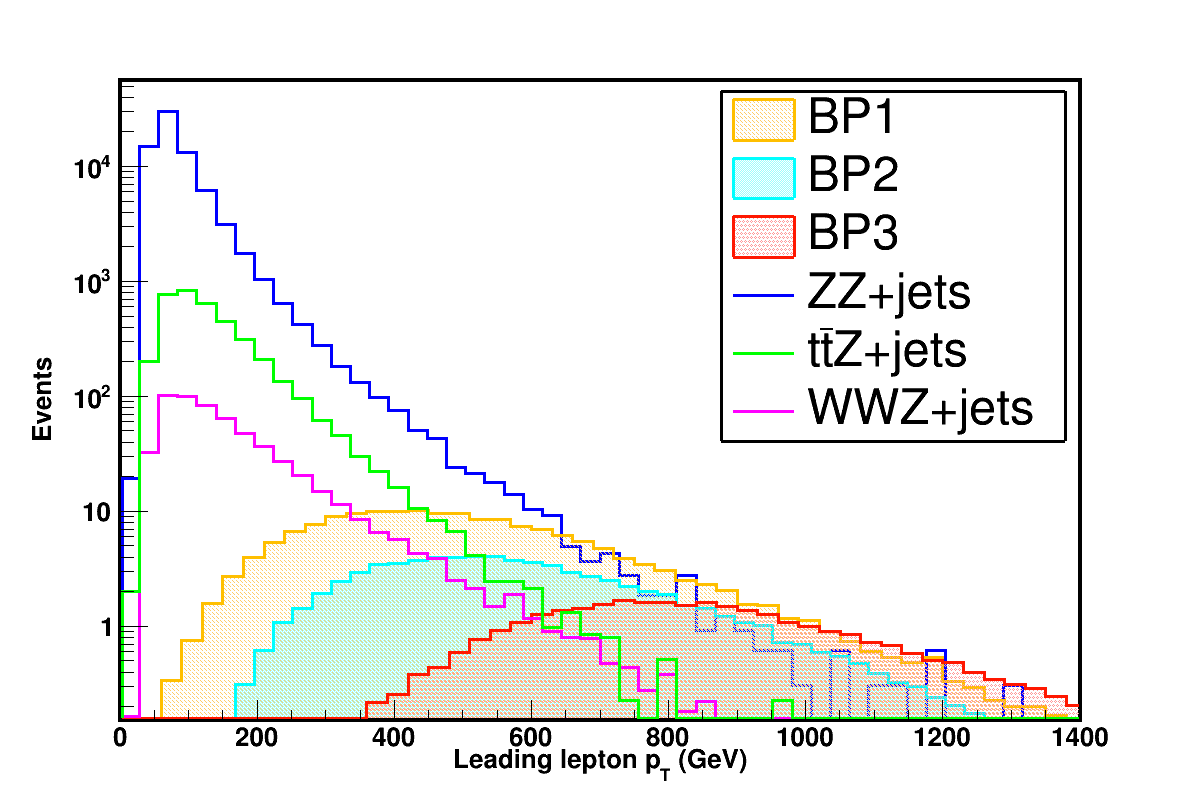}
   \caption{ Distributions of transverse momentum of leading lepton ($p_T^{l_1}$) at the HL-LHC 
   ($\sqrt{s}=14$ TeV with $\mathcal{L}=3000$ $fb^{-1}$) are shown here.  
The blue, green and magenta color solid lines represent the most dominant $ZZ + jets$, $WWZ+jets$ and $t\bar{t}Z+jets$ backgrounds. 
Yellow, cyan and red filled regions correspond to the benchmark points - \texttt{BP1}, \texttt{BP2} and \texttt{BP3} respectively. }
   \label{fig:pt_l1_14}
   \end{center}
\end{figure}

As mentioned earlier, $ZZ+jets$, $WWZ+jets$ and $t\bar{t}Z+jets$ are the most dominant backgrounds. We present the transverse momentum distribution of the leading lepton ($p_T^{l_1}$) of these dominant SM background channels along with signal corresponding to the three chosen benchmark points in Fig.~\ref{fig:pt_l1_14}. 
The blue, green and magenta color solid lines represent the distributions corresponding to the $ZZ + jets$, $WWZ+jets$ and $t\bar{t}Z+jets$ background channels respectively. The same for the benchmark points - \texttt{BP1}, \texttt{BP2} and \texttt{BP3} are shown in yellow, cyan and red filled regions, respectively. It is evident from the Fig.~\ref{fig:pt_l1_14} that for all the SM backgrounds the leading lepton $p_T$ peak occurs at a lower value compared to that for signals. Also, among the benchmark points, \texttt{BP3} has the largest $\mlspone$ which leads to shift of the peak to the higher value of $p_T$ compared to \texttt{BP1} and  \texttt{BP2}.  For both the signal regions (\texttt{SR-A} and \texttt{SR-B}), we choose a cut $p_T^{l_1} > 100$ GeV which will effectively discard the events coming from the SM backgrounds. To save computation time, therefore, $p_T^{l_1} > 100$ GeV cut is applied for all the background channels at the generation level itself. We have summarized the yield of background events after applying this generation level cuts in the last column of Table~\ref{tab:cs_bkg_14} in Appendix-\ref{appendix1}. This cut reduces the most dominant ZZ+jets background contribution by a factor of $\sim$5 whereas the signal events 
for benchmark points almost remain the same (reduced by only 1-2\%).

\begin{figure}[!htb] 
\begin{center}
       \includegraphics[width=0.8\textwidth]{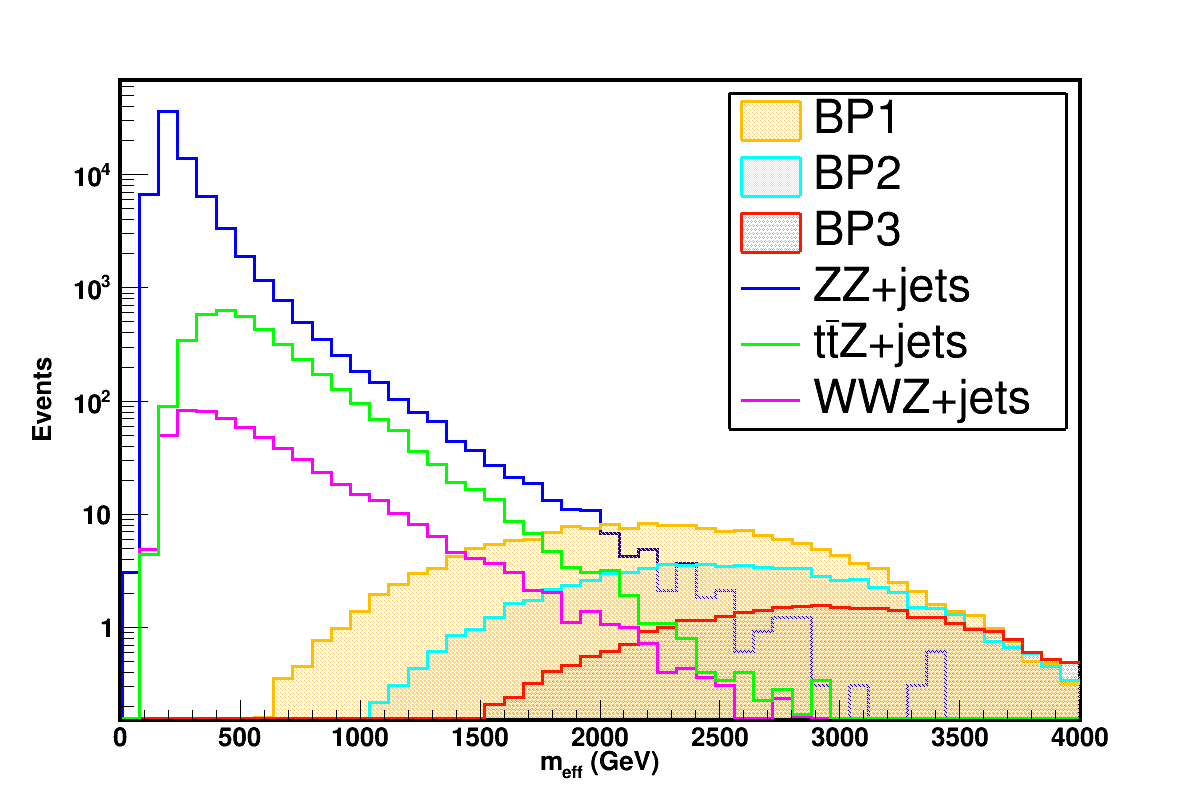}
    \caption{ Distributions of  effective mass ($m_{eff}$) at the HL-LHC ($\sqrt{s}=14$ TeV, 
        $\mathcal{L}=3000$ $fb^{-1}$) are shown here.  Color conventions are same as in Fig~\ref{fig:pt_l1_14}.}
    \label{fig:meff_14}
       \end{center}
\end{figure}

\begin{table}[!htb]
\begin{center}
\begin{tabular}{||c|c|c|c|c|c||}
\hline
\hline
\multirow{2}{*}{Cut variables}  & \multirow{2}{*} {} & \multirow{2}{*}{} & \multirow{2}{*}{} & \multicolumn{2}{c|}{Signal Region} \\ \cline{5-6}
   & \shortstack{$N_l\ge4$  \\ $(l=e,\mu)$ + \\$p_T^{l_1} >$ 100 GeV}  & Z veto & b veto & {\shortstack{SR-A \\($m_{eff}>900$)}} & {\shortstack{SR-B \\($m_{eff}>1500$)}}\\ 
 \hline 
 \hline
\texttt{BP1}   & 172.35 & 145.978 & 96.224 & 94.7357 & 81.346 \\ \hline
\texttt{BP2}  & 74.677 & 70.612 & 46.336 & 46.25 & 43.76 \\ \hline
\texttt{BP3}  & 32.422 & 30.834 & 19.559 & 19.55 & 19.289 \\ \hline
\hline
$ZZ$ + jets  & 17350  & 126.56 & 115.63 & 5.79 & 1.12 \\
$t\bar{t}Z$ + jets  & 2320  & 183.21 & 43.25 & 5.25 & 0.73 \\
$WWZ$ + jets  &378.77 & 29 & 25.67 & 6.32 & 1.33 \\
$WZZ$ + jets  & 217.78 & 3.83 & 3.19 & 0.71 & 0.13 \\
$ZZZ$ + jets  &104.76 & 1.02 & 0.78 & 0.07 & 0.01 \\
$h$ via GGF  & 1660 & 14.48 & 12.98 & 1.31 & 0.15 \\
$hjj$ & 66.84 & 17.7 & 14.63 & 0.51 & 0.01 \\
$Wh$ + jets  & 16.09 & 4.86 & 4.14 & 0.15 & 0.01 \\
$Zh$ + jets & 9.93 & 2.16 & 1.65 & 0.08 & 0.007 \\
\hline
\hline
\multicolumn{4}{||c|}{Total background} & 20.19 & 3.498 \\
\hline
\hline
\multicolumn{2}{||c|}{\multirow{3}{*}{\shortstack{Signal Significance $\sigma_{ss}$ \\ ( $\sigma_{ss}^{\epsilon}$, Sys. Unc.=5\%)}}}   &\multicolumn{2}{c|}{\texttt{BP1}} & 8.84 (7.79) & 8.83 (8.02) \\ \cline{3-6}
\multicolumn{2}{||c|}{}& \multicolumn{2}{c|}{\texttt{BP2}} & 5.67 (5.25) & 6.36 (6.02)\\ \cline{3-6} 
\multicolumn{2}{||c|}{}& \multicolumn{2}{c|}{\texttt{BP3}}  & 3.10 (2.96) & 4.04 (3.93) \\ \cline{3-6}
\hline
\hline
\end{tabular}
\caption{Selection cuts and the corresponding yields for the three signal benchmark points and relevant background channels at the HL-LHC are shown here. Statistical signal significance  ($\sigma_{ss}$) 
without any systematic uncertainty for \texttt{BP1}, \texttt{BP2} and \texttt{BP3} are also shown. Corresponding signal significance $\sigma_{ss}^{\epsilon}$ with Sys. Unc. $\epsilon$ = 5\% are presented in parenthesis. Here the SUSY signals belong to \texttt{Scenario-I.}}
\label{tab:cut_flow_14}
\end{center}
\end{table}

Among the relevant kinematic observables, $m_{eff}=\sum_i p_T^{l_i} + \sum_i p_T^{j_i} + \met$ turns out to be the most effective one. 
We observe that the maximum signal significance is  obtained by optimizing 
the $m_{eff}$ variable along with Z veto\footnote{Invariant mass of same-flavor-opposite-sign charged lepton pairs has to fall outside the window $101.2 \geq m_{ll} \geq 81.2$ GeV.} and a b-jet veto on the $N_l \ge 4$ final states. 
The $m_{eff}$ distributions for signal benchmark points and dominant backgrounds are 
depicted in Fig.~\ref{fig:meff_14}, where we have followed the same color conventions to represent signal benchmark points and background channels as Fig.~\ref{fig:pt_l1_14}. Similar to Fig.~\ref{fig:pt_l1_14}, the $m_{eff}$ distributions for SM background channels peak at much lower values compared to SUSY signals. The distributions corresponding to the signal benchmark points also differ depending on the choices of LSP and NLSP masses. Consequently, two signal regions are defined; \texttt{SR-A} and \texttt{SR-B} with 
$m_{eff} > 900$ GeV and  $m_{eff} > 1500$ GeV respectively. The $m_{eff}$ cut reduces the number of background events significantly. Apart from this, the Z-veto cut is most effective to reduce the $ZZ+jets$ events while b-veto is very effective to reduce $t\bar{t}Z+jets$ events.

Finally we estimate the statistical signal significance ($\sigma_{ss}$) using the relation  $\sigma_{ss}$ = ${S}/{\sqrt{S+B}}$ where $S$ and $B$ represent the signal and  background yield. The effects of systematic uncertainties are also shown by considering the formula $\sigma_{ss}^{\epsilon}$ = ${S}/{\sqrt{S+B+((S+B)\epsilon)^2}}$, where $\epsilon$ corresponds to the systematic uncertainty (Sys. Unc.).  
The list of selection cuts used for this cut-and-count analysis along with the  yield of signal benchmark points and the SM backgrounds after each cut for the HL-LHC are tabulated in Table~\ref{tab:cut_flow_14}. The signal significance $\sigma_{ss}$ without any systematic uncertainty for \texttt{BP1}, \texttt{BP2} and \texttt{BP3} are also shown in the last three rows of Table~\ref{tab:cut_flow_14}. We obtain that  $\sigma_{ss}$ for \texttt{BP1}, \texttt{BP2} and \texttt{BP3} are 8.84 (8.83), 5.67(6.36) and 3.1(4.04) respectively for the signal region \texttt{SR-A} (\texttt{SR-B}). \texttt{SR-B} consists of larger  $m_{eff}$ criteria compare  to \texttt{SR-A}  and is more effective to probe the parameter space with large $\mchonepm$ as evident in Table~\ref{tab:cut_flow_14}. We also present the  signal significance $\sigma_{ss} ^{\epsilon}$ with systematic uncertainty $\epsilon$ = 5\% for the benchmark points in 
Table~\ref{tab:cut_flow_14}. 
For \texttt{BP1}, \texttt{BP2} and \texttt{BP3}, the signal to background ratio (S/B) is $\sim$ 23, 12, 5 respectively for \texttt{SR-B}\footnote{for \texttt{SR-B}, the corresponding ratio  is $\sim$ 5, 2, 1 respectively.} and the 
\begin{figure}[!htb]
\begin{center}
\input{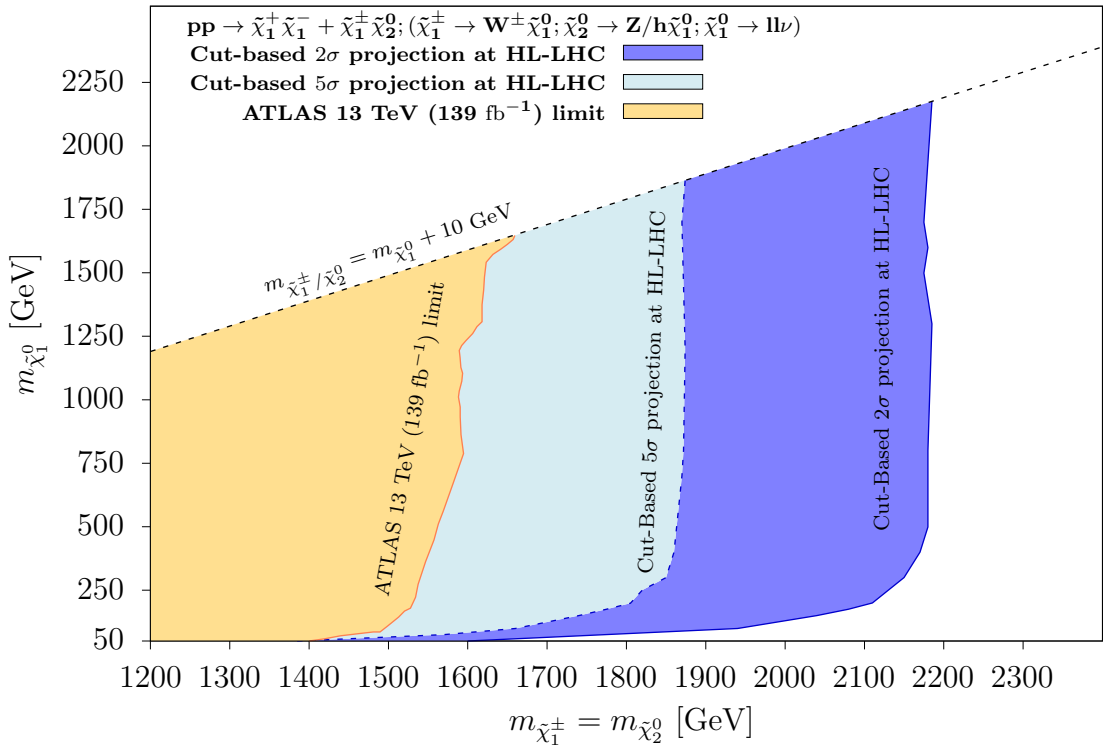}
\caption{ Projected discovery ($5\sigma$) and exclusion ($2\sigma$) regions in the $\mlsptwo-\mlspone$ mass plane at the HL-LHC are presented with light and dark blue colors. The yellow region represents the existing limit obtained by the ATLAS collaboration from Run-II data\cite{ATLAS:2021yyr}}.
\label{fig:reach_14tev_cut}
\end{center}
\end{figure}
changes in $\sigma_{ss}$ for including  the systematic uncertainty is not significant due to this large $S/B$ ratio.  We find that  $\epsilon$ = 5\%  
reduces the $\sigma_{ss}$ by $3 - 12$\%.

In Fig.~\ref{fig:reach_14tev_cut} we showcase the projected discovery region (with $\sigma_{ss}\geq 5$) and exclusion region (with $\sigma_{ss} \geq 2$) in the LSP-NLSP mass plane from direct $\chonepm \lsptwo$ + $\chonep \chonem $ production for \texttt{Scenario-I} at the HL-LHC. The projected 5$\sigma$ discovery and 2$\sigma$ exclusion regions obtained by traditional cut-and-count analysis are represented by the boundaries of light and dark blue colored regions respectively. The yellow region corresponds to the current 95\% C.L. observed limit obtained by the ATLAS collaboration using Run-II (13 TeV) 139 $fb^{-1}$  data \cite{ATLAS:2021yyr}. The black dashed line in Fig.~\ref{fig:reach_14tev_cut},  
represents the NLSP-LSP mass relation $\mlsptwo -\mlspone$ = 10 GeV. It is evident that HL-LHC is capable of extending the 5$\sigma$ projected discovery reach by around 200 GeV compared to the current LHC limit. Using this traditional cut-based analysis we also observe that the 95\% C.L. projected exclusion limits on  $\mlsptwo$=$\mchonepm$ reaches upto 2100 (2180) GeV at the HL-LHC for $\mlspone >$ 200 (500) GeV.

\begin{table}[!htb]
\begin{center}
\begin{tabular}{|c|c|c|c|c|} 
    \hline
	Benchmark & \multicolumn{4}{c|}{Signal Significance (Syst. Unc. = 5\%)} \\
	\cline{2-5}
	points & Scenario-I & Scenario-II & Scenario-III & Scenario-IV \\
	\hline
    \texttt{BP1} & 8.83 (8.02) & 6.46 (6.10) & 3.64 (3.56) & 2.33 (2.30)\\
    \hline
    \texttt{BP2} & 6.36 (6.02) & 4.67 (4.51) & 2.75 (2.70) & 1.69 (1.67) \\
    \hline
    \texttt{BP3} & 4.04 (3.93) & 2.93 (2.88) & 1.66 (1.64) & 0.92 (0.91) \\
    \hline\hline
    $m_{\lspone}$  & \multicolumn{4}{c|}{Projected exclusion on $m_{\chonepm}$ at the HL-LHC (Sys. Unc.= 20\%) } \\
	\hline
    800 & 2180 (2120) & 2080 (2020) & 1900 (1840) & 1740 (1680)\\
    \hline
\end{tabular}
\caption{ Comparison of signal significance of benchmark points \texttt{BP1}, \texttt{BP2} and \texttt{BP3} for different model scenarios (defined in Table~\ref{tab:br2}) with  0\% (5\%) systematic uncertainty are shown here. The numbers in last row represents the projected 95\% C.L. 2$\sigma$ exclusion limits on NLSP masses for a fixed 800 GeV LSP with 0\% (20\%) systematic uncertainty. Here all the masses are in GeV. }
\label{tab:scenario_compare}
\end{center}
\end{table}

We have defined four different scenarios (\texttt{Scenario-I,II,III,IV}) in Sec.~\ref{sec:model} (see Table~\ref{tab:br2}) obtained from different single non zero $\lambda_{ijk}$ couplings. The LSP pair gives $4l~(l = e, \mu) + \met$ final states  with 100\% branching ratios in \texttt{Scenario-I}. In this section, we have already discussed the prospect of \texttt{Scenario-I} for various benchmark points along with the projected exclusion in mass planes in great detail. For other scenarios, $\tau$ lepton appears in the final state with \texttt{Scenario-IV} being the most $\tau$ enriched. Now we proceed to explore the prospect of our conventional cut based $4l~(l = e, \mu)$ analysis for these other scenarios considering the already selected signal benchmark points. We present the signal significance of \texttt{BP1}, \texttt{BP2} and \texttt{BP3} for the four scenarios in Table~\ref{tab:scenario_compare} without and with a systematic uncertainty of $\epsilon$ = 5\%. As expected, the $\sigma_{ss}$ is maximum for \texttt{Scenario-I} and minimum for \texttt{Scenario-IV} owing to the comparatively lower $\tau$ tagging efficiency compared to other charged leptons. As Fig.~\ref{fig:reach_14tev_cut} shows, the exclusion limits on $\mchonepm$ are almost constant for a relatively larger $\mlspone$, we 
also estimate the projected exclusion limits on $m_{\chonepm}$ at the HL-LHC for the four models by fixing the LSP mass at 800 GeV (same with  \texttt{BP2}). We derive that for this specific choice of $\mlspone$, the projected $2\sigma$ exclusion limits on $\mchonepm (=\mlsptwo)$ become 2.18, 2.08, 1.90 and 1.74 TeV for \texttt{Scenario-I}, \texttt{Scenario-II}, \texttt{Scenario-III}, and \texttt{Scenario-IV} respectively. It is also observed that the mass limits gets reduced by $\sim$ 60 GeV for systematic uncertainty $\epsilon$ = 20\% (refer to the last row in Table~\ref{tab:scenario_compare}).

\subsection{Prospect at the HL-LHC using Machine Learning based analysis}
\label{sec:14_ml}

We now proceed to use a boosted decision tree ({\tt BDT}) based machine learning algorithm to assess if the results of our cut-based analysis can be improved upon. For this purpose, we construct the following set of 18 kinematical variables (also called `features' in ML language) taking into account the kinematics of the multilepton final state. 
\begin{itemize}
\item Transverse momenta of leading lepton ($p_T^{l_1}$) and subleading lepton ($p_T^{l_2}$)~\footnote{The ordering of leptons are based on their transverse momenta, $l_1$ denotes the lepton with highest $p_T$ at each event, $l_2$  denotes the lepton with second highest $p_T$ value at that event and so on.} (2 variables).
\item $\Delta R$ ($\Delta R = \sqrt{(\Delta \eta)^2 + (\Delta \phi)^2}$, where $\eta$ is pseudo-rapidity and $\phi$ is azimuthal angle) between leading lepton and other subleading leptons, denoted as $\Delta R_{l_1l_2}$, $\Delta R_{l_1l_3}$, $\Delta R_{l_1l_4}$; and similarly $\Delta R$ between other leptons, $\Delta R_{l_2l_3}$, $\Delta R_{l_2l_4}$ and $\Delta R_{l_3l_4}$ (6 variables)
\item Difference in azimuthal angle between leptons and missing transverse momenta, which are $\Delta \phi_{l_1 \met}$, $\Delta \phi_{l_2 \met}$, $\Delta \phi_{l_3 \met}$ and $\Delta \phi_{l_4 \met}$ (4 variables)
\item Number of jets at each event, both b-tagged jets ($N_b$) and non b-tagged jets ($N_j$) (2 variables)
\item Missing transverse energy ($\met$) and effective mass ($m_{eff}$) as defined in Sec.~\ref{sec:14_cut} (2 variables)
\item Number of Same Flavor Opposite Sign lepton pair ($N_{SFOS}$) and Number of SFOS pair lies within the range $81.2\leq m_{SFOS}\leq 101.2$ GeV ($N_Z$) (2 variables)
\end{itemize}

For our ML-based  multi-variate analysis we have used Extreme Gradient Boosted decision tree algorithm through {\tt XGBoost} machine learning toolkit \cite{Chen:2016btl}. Training and testing of the XGBoost module are done by implementing the multiclass classification through {\tt multi:softprob} objective function. The SUSY signal events and the SM backgrounds 
events which contain atleast 4$l (l= e, \mu)$ are only considered~\footnote{No generation level $p_T$ cuts are applied for the signal/background events.}. Details of lepton and jets identification, isolation criteria etc. are already summarized in Sec.\ref{sec:14_cut}. 
After calculating the 18 kinematic observables, the signal and background events are mixed with proper weight according to their 
relative cross-sections. We have used 80\% of this data set for training and the remaining for testing. The hyperparameters {\tt learning rate, number of trees}, {\tt maximum depth  } are tuned to optimize the signal significance.  The {\tt number of trees} and  
{\tt maximum depth  } of a tree are chosen as 500 and 10 respectively. The  {\tt learning rate ($\eta$) } parameter or step size shrinkage is chosen in the range [0.01-0.03] to prevent over-fitting. The $\eta$ parameter shrinks the features weights at each boosting 
step which makes the process more conservative \cite{xgb_param}. The {\tt multi:softprob} object function returns the predicted probability score of each data point belonging to each class (signal and multiple backgrounds). To obtain the discovery and exclusion contours on the gaugino mass plane we have applied a threshold on the probability score to obtain maximum significance.

\begin{figure}[!htb]
\begin{center}
\includegraphics[width=0.49\textwidth]{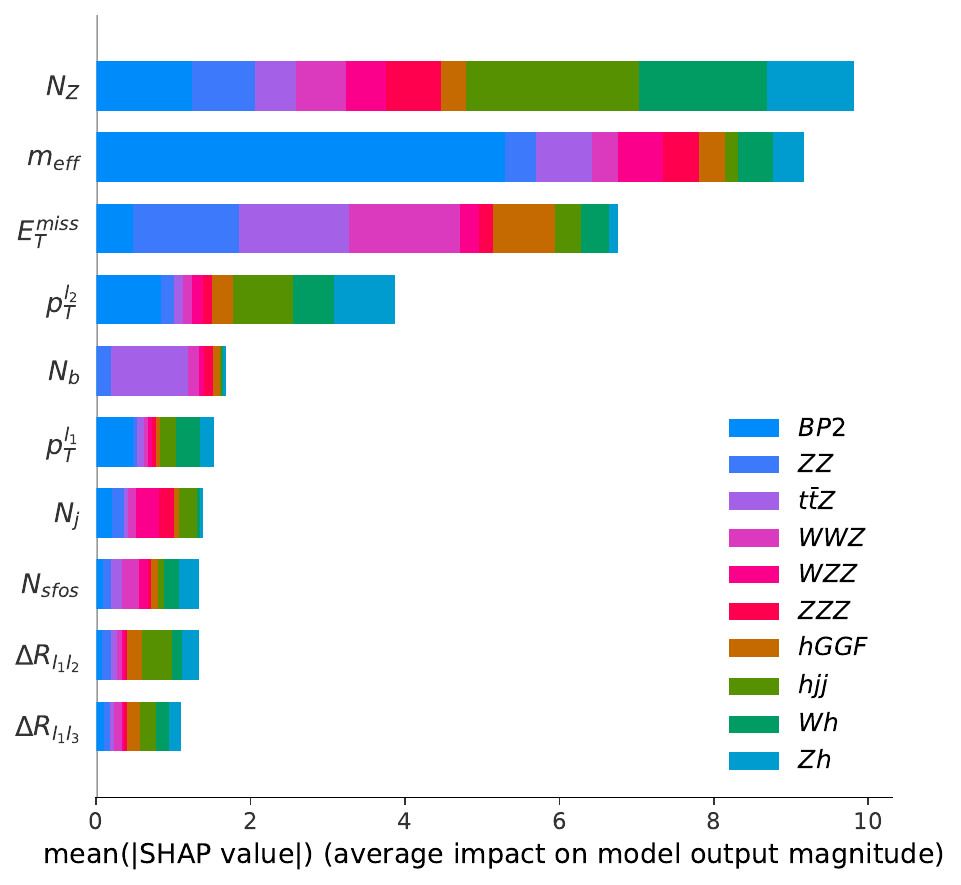}
\includegraphics[width=0.5\textwidth]{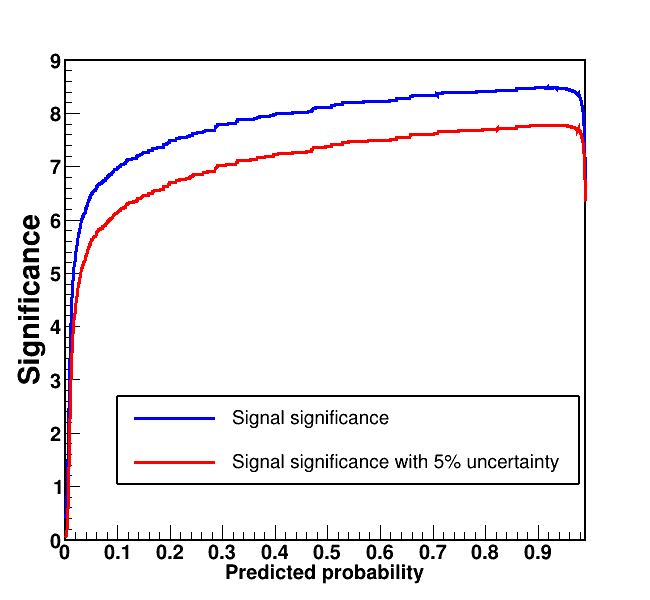}
\caption{(Left) Shapley feature importance plot for the top 10 important kinematic variables for data set with benchmark SUSY signal  BP2 ($\mchonepm =$1800, $\mlspone=$800) and backgrounds analyzed at HL-LHC. (Right) The signal significance without any systematic uncertainty (blue line) and with 5\% systematic uncertainty (red line) as a function of predicted probability are shown. }
\label{shap_scoreBP2}
\end{center}
\end{figure}

All the kinematic variables  or the  `features' are  not equally effective in predicting the signal and different background classes. To understand the effect of each feature in predicting each class distinctly, we computed Shapley values using SHapley Additive ex-Planations (SHAP) \cite{DBLP:journals/corr/LundbergL17,Shapley+1953+307+318} package. In collider studies, SHAP values are very helpful to understand the effect of each feature on the model's output \cite{DBLP:journals/corr/abs-1802-03888,Grojean:2020ech,Alvestad:2021sje,Cornell:2021gut}. To find out the average marginal contribution of a feature,  SHAP finds out the difference between the two outputs of model prediction by training the model with the feature and also excluding the same feature. Then it calculates the weighted average of the possible differences for different subsets of all features \cite{DBLP:journals/corr/abs-1802-03888}. In this way, the global feature importance is calculated for every feature  and furthermore, the SHAP value of each feature for every event and the mean of absolute Shapley values are calculated by averaging over all the events. According to the mean of the absolute values, SHAP ranks the features and the  feature importance plot for the top 10 kinematic variables for the benchmark SUSY signal  BP2 ($\mchonepm =$1800, $\mlspone=$800) and various backgrounds is displayed in the left panel of Fig.~\ref{shap_scoreBP2}. We observe that $N_Z$ has the most significant effect on predicting the signal and backgrounds for BP2 and the next 5 important features are  $m_{eff}$, $\met$, $p_T^{l_2}, N_b$ and $p_T^{l_1}$. For the cut-based analysis we  also find that these variables are effective to discriminate the SUSY signal and  backgrounds (refer to Fig.~\ref{fig:pt_l1_14}-\ref{fig:meff_14} for $p_T^{l_1}$ and $m_{eff}$  distribution, Table~\ref{tab:cut_flow_14} for cut flow and for more details refer to Sec.~\ref{sec:14_cut}). In Fig.~\ref{shap_scoreBP2} (left), the spread of the color bar along x-axis corresponds to the contribution of that feature to classify that particular class. 

\begin{table}[!htb]
\begin{center}
\begin{tabular}{|c|c|c|c|c|c|}
\hline
Benchmark & Probability & Signal & Total   & Signal  &  Gain in $\sigma_{ss}$ \\
Points & Score & Yield & Background &  Significance $\sigma_{ss}$ & from  \\
& &  & Yield &  (Sys Unc. = $5\%$) & Cut-based \\
\hline
\hline
\texttt{BP1} & 0.90  & 165.80 & 6.99  & 12.61 (10.54) & 43\% (31\%)\\
\cline{2-6}
(1600,250) & 0.96 & 156.89 & 3.94 &  12.37 (10.45) & 40\% (30\%) \\
\hline
\hline
\texttt{BP2} & 0.90  & 73.84  & 1.96  & 8.48 (7.78) & 33\% (29\%)  \\
\cline{2-6}
(1800,800) & 0.96 &  72.47 & 1.34 & 8.44 (7.75) & 33\% (29\%) \\
\hline
\hline
\texttt{BP3} & 0.90  & 32.32  & 1.32  & 5.57 (5.35) & 38\% (36\%)  \\
\cline{2-6}
(1950,1850) &  0.96 &  31.97 & 0.30 & 5.63 (5.41) & 40\% (38\%) \\
\hline
\hline
\end{tabular}
\caption{Signal yield, total background yield and the signal significance (without any systematic uncertainity) at the HL-LHC using ML-based algorithm for different probability scores are presented here. The numbers in the parenthesis correspond to $\sigma_{ss}$ with systematic uncertainty $\epsilon = 5\%$. Here the SUSY signal belongs to 
{\tt{Scenario-I.}}}
\label{tab:significance_14_ml}
\end{center}
\end{table}

\begin{figure}[!htb]
\begin{center}
\input{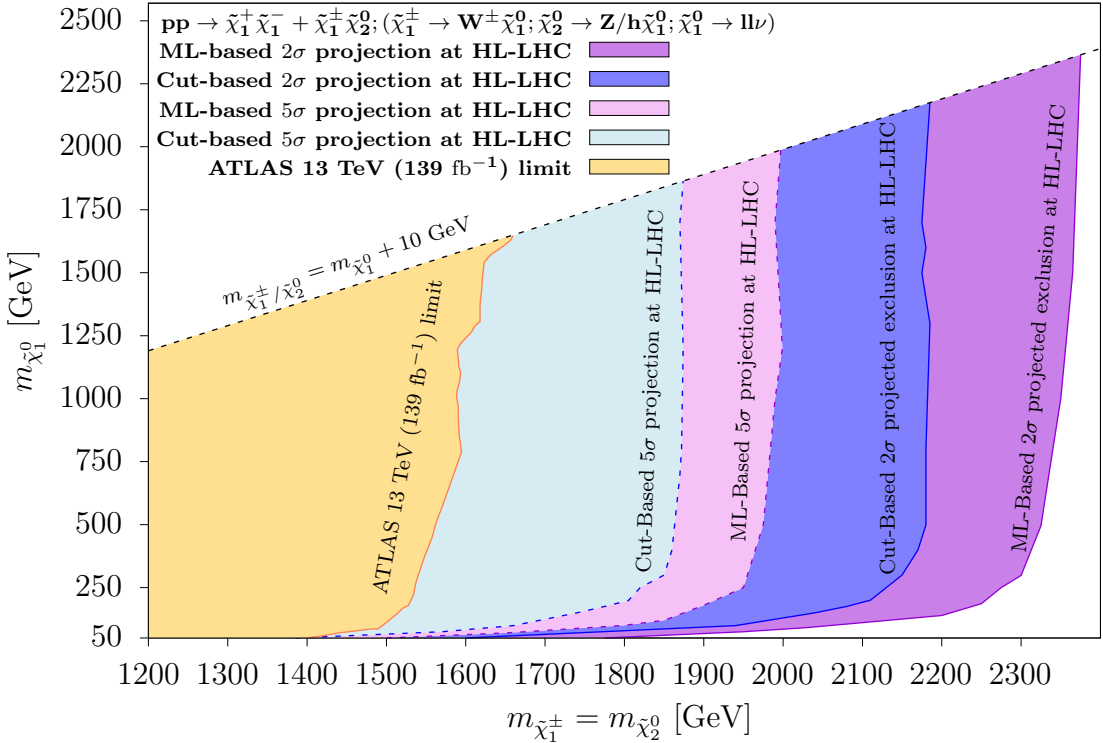}
\caption {Projected discovery ($5\sigma$) and exclusion ($2\sigma$) reach in the 
$\mlsptwo-\mlspone$ mass plane at the HL-LHC are presented with light and dark violet colors. 
For the light blue, dark blue and yellow color, the color conventions are same as in Fig.\ref{fig:reach_14tev_cut} The yellow regions represents the existing limit obtained by the ATLAS collaboration from Run-II data\cite{ATLAS:2021yyr} }
\label{fig:reach_14tev_ml}
\end{center}
\end{figure}

In the right panel of Fig.~\ref{shap_scoreBP2}, we present the variation of signal significance  as a function of probability score cut. The blue line corresponds to $\sigma_{ss}^{\epsilon}$ with systematic uncertainty $\epsilon$ = 0 \% and the red line represents the same with $\epsilon$ = 5\%. It is evident that the signal significance reaches a maximum and saturates around probability score $\sim$ 0.90 - 0.95. We present the signal yield, total background yield and the signal significance (without any systematic uncertainty) at the HL-LHC in the Table~\ref{tab:significance_14_ml} for two sample values of probability score cut 0.90 and 0.96 chosen from the saturated region. Comparing the signal significance obtained via the traditional cut-and-count method (refer to Table~\ref{tab:cut_flow_14}), we observe that roughly $\sim$ 30-40\% gain is achieved for ML-based analysis. Numbers in the parenthesis correspond to $\sigma_{ss}$ and gain with systematic uncertainty $\epsilon = 5\%$. It may be noted that similar to   Table~\ref{tab:cut_flow_14}, here the SUSY signal belongs to {\tt{Scenario-I}}.   For this same scenario, we also estimate the projected $5\sigma$ discovery reach   and $2\sigma$ exclusion regions in the $\mlsptwo-\mlspone$ mass plane at the HL-LHC in Fig.~\ref{fig:reach_14tev_ml}, represented by light and dark violet color respectively. The light and dark blue regions correspond to projected 5$\sigma$ and 2$\sigma$ regions obtained from the conventional cut-and-count method. Similar to Fig.~\ref{fig:reach_14tev_cut}, 
the yellow regions are already ruled out by the ATLAS 13 TeV data \cite{ATLAS:2021yyr}.

As evident from Table~\ref{tab:significance_14_ml}, the ML-based analysis improves the sensitivity by $\sim$ 30-40\%  due to its superior capability of segregating signal from various background channels. This leads to a greater reach for discovery and exclusion in the bino-wino mass plane (Fig.\ref{fig:reach_14tev_ml}). The projected discovery reach extends to $\sim 1.99$ TeV in ML-based methods with an enhancement of 120 GeV compare to cut-based method. Similarly, the projected exclusion curve reaches $\sim 2.37$ TeV from our cut based estimate of $\sim 2.18$ TeV (enhancement of 190 GeV). In Table~\ref{tab:scenario_compare_14_ml}, we have presented the signal significance corresponding to different benchmark points for each SUSY scenario (as defined in Table~\ref{tab:br2}). Similar to  cut-based analysis, the $\sigma_{ss}$ is maximum for \texttt{Scenario-I} and minimum for the \texttt{Scenario-IV}. We have also shown the projected exclusion limit found for each scenario at the LSP mass 800 GeV without and with 20\% systematic uncertainty. As similar to $\sigma_{ss}$, the $2\sigma$ reach is also maximum for \texttt{Scenario-I} (2340 GeV) and minimum for \texttt{Scenario-IV} (1935 GeV) and the gap between the two $2\sigma$ reaches corresponding to without and with 20\% systematic uncertainty is around 65 GeV for each scenario. It may be noted that  the 
projected exclusion limits on $\mchonepm$ as mentioned in Table~\ref{tab:scenario_compare_14_ml} improve by $\sim$ 30 GeV if the LSP-NLSP mass gap is 
$\sim$ 10 GeV.

\begin{table}[!htb]
\begin{center}
\begin{tabular}{|c|c|c|c|c|} 
    \hline
	Benchmark & \multicolumn{4}{c|}{Signal Significance (Syst. Unc. = 5\%)} \\
	\cline{2-5}
	points & Scenario-I & Scenario-II & Scenario-III & Scenario-IV \\
	\hline
    \texttt{BP1} & 12.61 (10.54) & 9.49 (8.49) & 5.95 (5.67) & 4.21 (4.1)\\
    \hline
    \texttt{BP2} & 8.48 (7.78) & 6.45 (6.11) & 4.14 (4.03) & 2.96 (2.92) \\
    \hline
    \texttt{BP3} & 5.57 (5.35) & 4.21 (4.11) & 2.85 (2.82) & 1.51 (1.5) \\
    \hline\hline
    $m_{\lspone}$  & \multicolumn{4}{c|}{Projected exclusion on $m_{\chonepm}$ at the HL-LHC (Sys. Unc.= 20\%) } \\
	\hline
    800 & 2340 (2275) & 2240(2175) & 2050 (1985) & 1935 (1870)\\
    \hline
\end{tabular}
\caption{Comparison of signal significance of benchmark points \texttt{BP1}, \texttt{BP2} and \texttt{BP3} for different model scenarios 
(defined in Table~\ref{tab:br2}) with  0\% (5\%) systematic uncertainty are shown here. The numbers in last row represents the projected 95\% C.L. 2$\sigma$ exclusion limits on NLSP masses for a fixed 800 GeV LSP with 0\% (20\%) systematic uncertainty. Here all the masses are in GeV. }
\label{tab:scenario_compare_14_ml}
\end{center}
\end{table}


\subsection{Prospect at the HE-LHC using cut-based analysis}
\label{sec:27_cut}
In this section, we present the cut-and-count analysis for the search of wino pair production
\begin{figure}[!htb]
\begin{center}
               \includegraphics[width=0.485\textwidth]{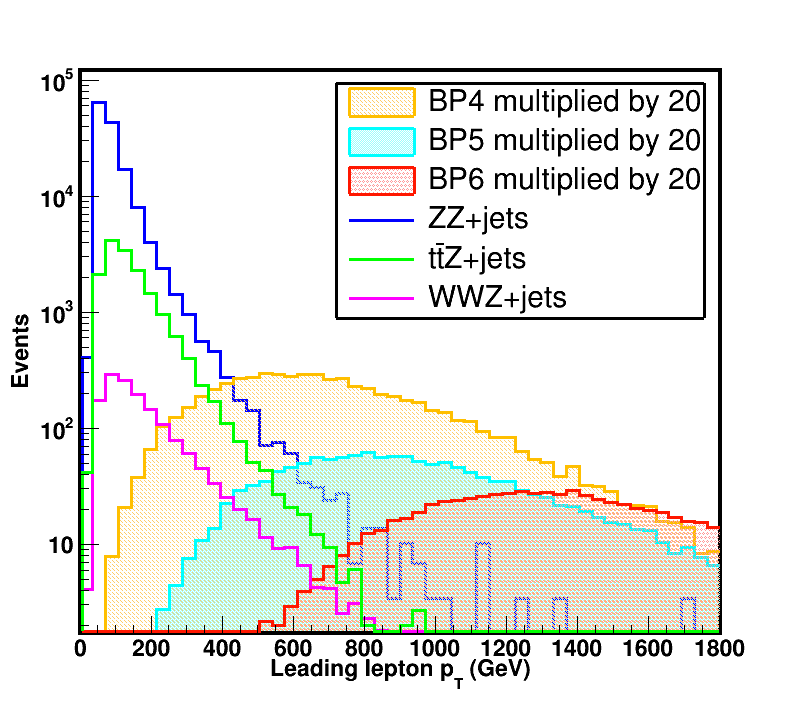}
           \includegraphics[width=0.485\textwidth]{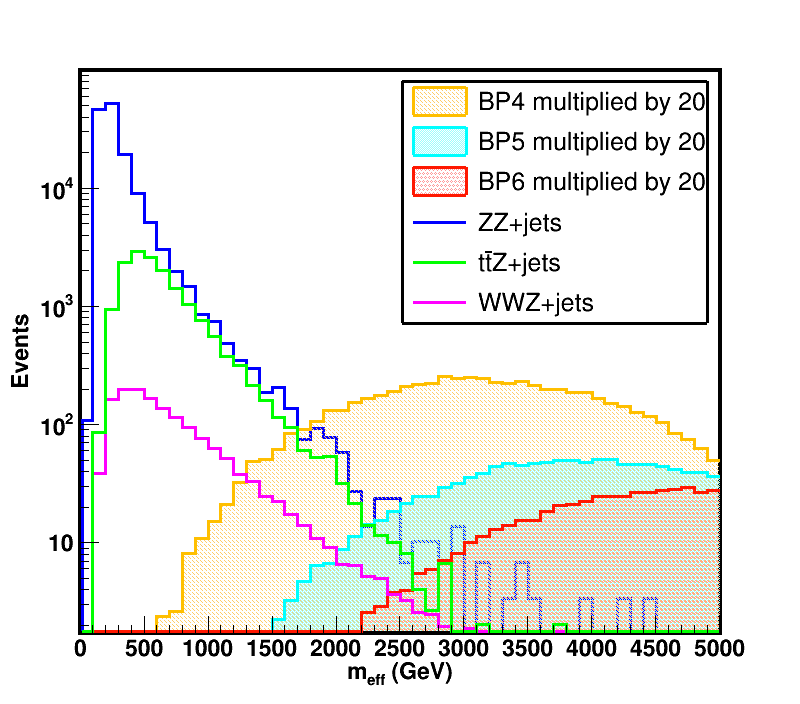}

   \caption{ Distributions of transverse momentum of leading lepton $p_T^{l_1}$ (left panel)  
   and effective mass $m_{eff}$ (right panel) at the HE-LHC. 
The blue, green and magenta color solid lines represent the most dominant $ZZ + jets$, $WWZ+jets$ and $t\bar{t}Z+jets$ backgrounds respectively. 
Yellow, cyan and red filled regions correspond to the benchmark points - \texttt{BP4}, \texttt{BP5} and \texttt{BP6} respectively. 
 }
   \label{fig:pt_l1_27}
   \end{center}
\end{figure}
at the High Energy LHC (HE-LHC) with $N_l \geq 4$ final state at $\sqrt{s} = 27$ TeV and $\mathcal{L} =$ 3000 $fb^{-1}$ as presented in Sec.~\ref{sec:14_cut} for HL-LHC. For this analysis, we have defined two signal regions as \texttt{SR-C} and \texttt{SR-D} with $m_{eff} > 1500$ GeV and $m_{eff} > 2200$ GeV respectively along with other cuts like $p_T^{l_1} >$ 150 GeV, Z-veto and b-veto. 
We have defined three representative signal benchmark points -
\texttt{BP4}: $\mchonepm = 2300$ GeV, $\mlspone = 500$ GeV, 
\texttt{BP5}: $\mchonepm = 2900$ GeV, $\mlspone = 1200$ GeV, 
\texttt{BP6}: $\mchonepm = 3100$ GeV, $\mlspone = 3000$ GeV.
The cross-sections of wino pair production corresponding to these benchmark points at NLO+NLL level are mentioned in Table~\ref{tab:cs_sig} in Appendix-\ref{appendix2}. 

The transverse momentum distribution of the leading lepton ($p_T^{l_1}$) and effective mass ($m_{eff}$) are shown in Fig.~\ref{fig:pt_l1_27} for the leading backgrounds with blue, green and magenta colored solid lines corresponding to $ZZ + jets$, $WWZ+jets$ and $t\bar{t}Z+jets$ respectively. The signal benchmark points  \texttt{BP4}, \texttt{BP5} and \texttt{BP6} are shown by yellow, cyan and red filled regions respectively. 

\begin{table}[h!]
\small
\centering
\begin{tabular}{||c|c|c|c|c|c||}
\hline\hline
\multirow{2}{*}{} & \multirow{2}{*} {} & \multirow{2}{*}{} & \multirow{2}{*}{} & \multicolumn{2}{c||}{Signal Region} \\ \cline{5-6}
Cut variables & \shortstack{$N_l\geq4$\\$(l=e,\mu)$ + \\$p_T^{l_1} >$ 150 GeV} & Z veto & b veto & \multirow{2}{*}{\shortstack{SR-C \\($m_{eff}>1500$)}} & \multirow{2}{*}{\shortstack{SR-D \\($m_{eff}>2200$)}}\\ 
& & & & &  \\ \hline \hline
BP4 (2300,250) & 307.61 & 266.84 & 179.46 & 173.43 & 147.48 \\ \hline
BP5 (2900,1200) & 71.72 & 69.89 & 47.51 & 47.31 & 45.54 \\ \hline
BP6 (3100,3000) & 41.19 & 39.57 & 25.06 & 24.97 & 24.77 \\
\hline
\hline
$ZZ$ + jets & 15980  & 125.38 & 108.31 & 6.01 & 1.2 \\
$t\bar{t}Z$ + jets & 5814  & 467.27 & 103.94 & 6.77 & 1.73 \\
$WWZ$ + jets & 742.03 & 57.42 & 47.49 & 8.21 & 2.29 \\
$WZZ$ + jets & 414.87 & 7.93 & 6.02 & 1.09 & 0.27 \\
$ZZZ$ + jets & 142.17 & 1.47 & 1.06 & 0.08 & 0.02 \\
$h$ via GGF & 3490 & 34.51 & 29.30 & 1.47 & 0.33 \\
$hjj$ & 40.59 & 9.92 & 7.86 & 0.07 & 0 \\
$Wh$ + jets & 9.81 & 3.04 & 2.53 & 0.03 & 0.003 \\
$Zh$ + jets & 7.08 & 1.42 & 1.06 & 0.02 & 0.003 \\
\hline\hline
\multicolumn{4}{||c|}{Total background} & 23.76 & 5.86 \\
\hline
\hline
\multicolumn{2}{||c|}{\multirow{3}{*}{\shortstack{Signal Significance $\sigma_{ss}$ \\ ($\sigma_{ss}^{\epsilon}$, Syst. Unc. = 5 \%)}}}   &\multicolumn{2}{c|}{\texttt{BP4}} & 12.35 (10.10) & 11.90 (10.12) \\ \cline{3-6}
\multicolumn{2}{||c|}{}& \multicolumn{2}{c|}{\texttt{BP5}} & 5.61 (5.17) & 6.35 (5.98)\\ \cline{3-6} 
\multicolumn{2}{||c|}{}& \multicolumn{2}{c|}{\texttt{BP6}}  & 3.58 (3.37) & 4.47 (4.31)  \\ \cline{3-6}
\hline\hline

\end{tabular}
\captionof{table}{Selection cuts and the corresponding yields for three signal benchmark points and background channels at HE-LHC with ${\mathcal L}=3000~fb^{-1}$ are shown here.
Statistical signal significance  ($\sigma_{ss}$) 
without any systematic uncertainty for \texttt{BP4}, \texttt{BP5} and \texttt{BP6} are also shown. Corresponding signal significance $\sigma_{ss}^{\epsilon}$ with Sys. Unc. $\epsilon$ = 5\% are presented in parenthesis. Here the SUSY signals belong to \texttt{Scenario-I.}
}
\label{tab:cut_flow_27}
\end{table}

Similar to HL-LHC analysis, from the Fig.~\ref{fig:pt_l1_27}, we can see that the distributions corresponding to the SM backgrounds peak at lower $p_T$ values and the signal distributions are considerably more spread out peaking at much higher $p_T$ values. Consequently, we have chosen $p_T^{l_1} > 150$ cut at the generation level to generate the SM background processes in order to save computation time. The cross-sections and the yields after the $p_T$ cut of all the SM background processes are summarized in Table~\ref{tab:cs_bkg_27} of Appendix-\ref{appendix3}. From the distribution of $m_{eff}$ variable shown in Fig.~\ref{fig:pt_l1_27}, it is evident that the distributions for the SM processes peak at a much lower value of $m_{eff}$ compared to those of the signal benchmark points. By optimizing the signal regions for different kinematic variables, we have found that similar to the 14 TeV analysis, a combination of $m_{eff}$ variable with Z veto and b-jet veto maximize the signal significance for the $N_l \geq 4$ final state. The cut-flow table for the signal benchmark points and the SM background channels are summarized in the Table~\ref{tab:cut_flow_27}. 
\begin{figure}[!htb]
\centering
\input{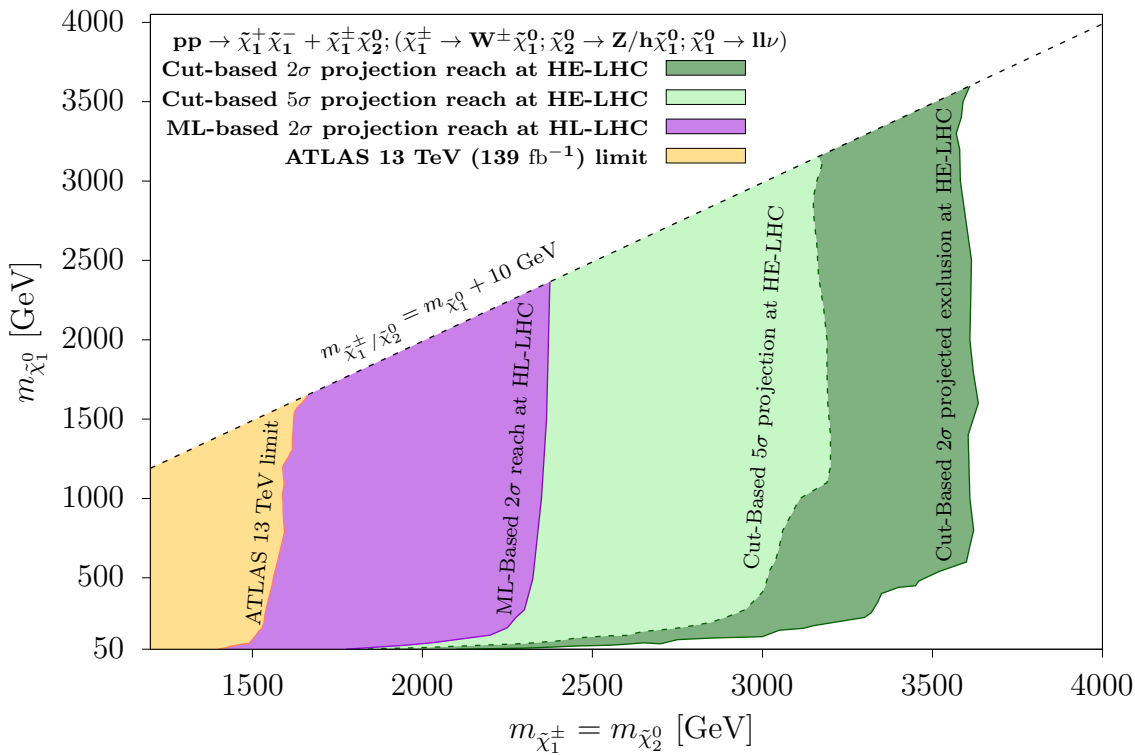}
\caption{ Projected discovery ($5\sigma$) and exclusion ($2\sigma$) regions in the $\mlsptwo-\mlspone$ mass plane at the HE-LHC via conventional cut-and-count method are presented with light and dark green colors. Also, the dark violet color region corresponds to 2$\sigma$ reach at HL-LHC obtained by ML-based analysis which is already summarized in Sec~\ref{sec:14_ml} . The yellow regions represents the existing limit obtained by the ATLAS collaboration from Run-II data \cite{ATLAS:2021yyr}.}
\label{fig:exclusion_plot_27_cut} 
\end{figure}
The signal significances for the benchmark points with and without a $5\%$ uncertainty are also shown in the Table~\ref{tab:cut_flow_27} for our two signal regions. We have obtained $\sigma_{ss}=$ 12.35 (11.90), 5.61 (6.35) and 3.58 (4.47) for BP4, BP5 and BP6  respectively for \texttt{SR-C} (\texttt{SR-D}). It is evident that \texttt{SR-D} signal region is more effective for the higher $\mchonepm$ mass. The signal to background yield ratio (S/B) at 27 TeV of BP4, BP5 and BP6 are $\sim$ 25, (14), 8 (2), 4 (1) respectively for \texttt{SR-D} (\texttt{SR-C}). Because of the large S/B ratio for \texttt{SR-D}, the changes in $\sigma_{ss}$ is less significant when systematic uncertainty is considered to be non-zero. For \texttt{SR-D}, the signal significance ($\sigma_{ss}$) reduces by $3 - 15$\% in presence of 5\% systematic uncertainty. 

\begin{table}[tb!]
\begin{center}
\begin{tabular}{|c|c|c|c|c|} 
    \hline
	Benchmark & \multicolumn{4}{c|}{Cut based Signal Significance (Syst. Unc. = 5\%)} \\
	\cline{2-5}
	points & Scenario-I & Scenario-II & Scenario-III & Scenario-IV \\
	\hline
    \texttt{BP4} & 11.90 (10.12) & 8.67 (7.86) & 5.96 (5.64) & 3.02 (2.96)\\
    \hline
    \texttt{BP5} & 6.35 (5.98) & 4.67 (4.49) & 2.77 (2.71) & 1.66 (1.64) \\
    \hline
    \texttt{BP6} & 4.47 (4.31) & 3.23 (3.15) & 1.83 (1.80) & 1.06 (1.05) \\
    \hline\hline
    $m_{\lspone}$  & \multicolumn{4}{c|}{Projected exclusion on $m_{\chonepm}$ at the HE-LHC (Sys. Unc.= 20\%) } \\
	\hline
    1200 & 3620 (3480) & 3400 (3260) & 3080 (2940) & 2780 (2640)\\
    \hline
\end{tabular}
\caption{Comparison of signal significance of benchmark points \texttt{BP4}, \texttt{BP5} and \texttt{BP6} for different model scenarios 
(defined in Table~\ref{tab:br2}) with  0\% (5\%) systematic uncertainty are shown here. The numbers in last row represents the projected 95\% C.L. 2$\sigma$ exclusion limits on NLSP masses for a fixed 1200 GeV LSP with 0\% (20\%) systematic uncertainty. Here all the masses are in GeV.  }
\label{tab:scenario_compare_27}
\end{center}
\end{table}

For HE-LHC, we now show the projected discovery region (with $\sigma_{ss} \geq 5$) and exclusion region (with $\sigma_{ss} \geq 2$) in Fig.~\ref{fig:exclusion_plot_27_cut}   with light and dark green colors respectively for \texttt{Scenario-I}. The dark violet color region corresponds to 2$\sigma$  projection at HL-LHC obtained by ML-based analysis which is already displayed in Fig.~\ref{fig:reach_14tev_ml} in Sec.~\ref{sec:14_ml} and the yellow region is the current limits obtained by the ATLAS collaboration using Run-II data \cite{ATLAS:2021yyr}. We find that 95\% $C.L.$ projected exclusion limits on $\mlsptwo$=$\mchonepm$ reaches upto $\sim$ 3.5 (3.6) TeV at the HE-LHC for $\lspone >$ 500 (1000) GeV. We also obtain that the 5$\sigma$ projected discovery reach will be around 3.16 TeV.
Now we proceed to compare our results corresponding to \texttt{SR-D} for the four different scenarios (\texttt{Scenario-I, II, III, IV}) defined in Table~\ref{tab:br2} in Sec.~\ref{sec:model}. The results are summarized in Table~\ref{tab:scenario_compare_27}. As in the 14 TeV case, the signal significance steadily decreases with increasing $\tau$ lepton multiplicity in the final state with \texttt{Scenario IV} being the least sensitive of all. $\sigma_{ss}$ are quoted for the three benchmark points with systematic uncertainty $0\%$ ($5\%$). 
The last row represents the projected $2\sigma$ exclusion limit on the wino masses at 95\% C.L. keeping the LSP mass fixed at 1200 GeV with 0\% (20\%) systematic uncertainty. 
We find that the projected $2\sigma$ limit will be $\sim$ 800 GeV weaker for 
\texttt{Scenario-IV} compare to the other extreme model i.e., \texttt{Scenario-I}. 


\subsection{Prospect at the HE-LHC using Machine Learning based analysis}
\label{sec:27_ml}

\begin{figure}[!htb]
\begin{center}
\includegraphics[width=0.49\textwidth]{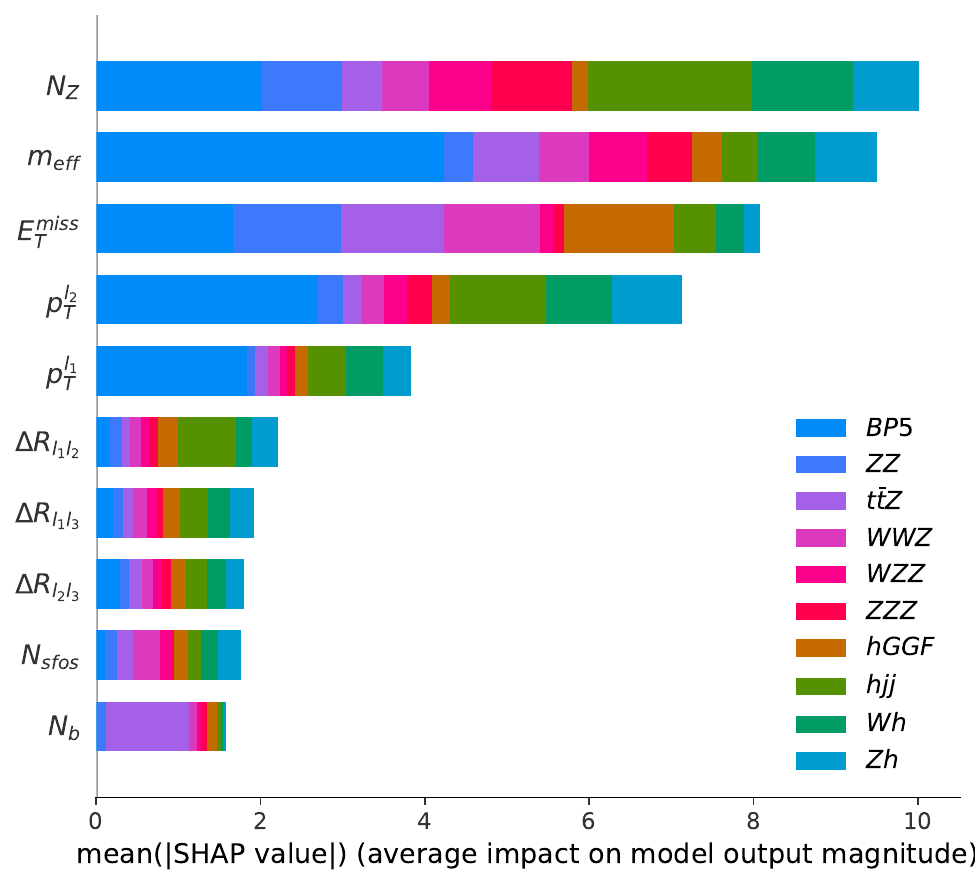}
\includegraphics[width=0.5\textwidth]{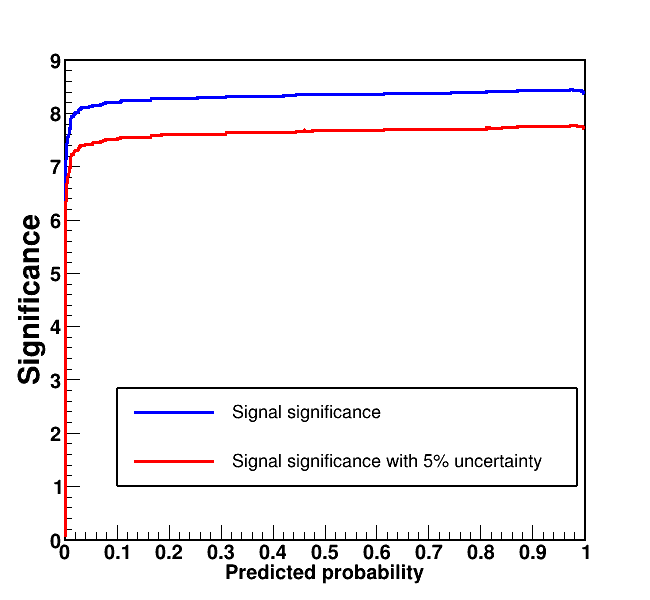}
\caption{ (Left) Shapley feature importance plot for the top 10 kinematic variables for data set with benchmark SUSY signal  BP5 ($\mchonepm =$2900, $\mlspone=$1200) and backgrounds analyzed at HE-LHC. (Right) The signal significance for BP5 without any systematic uncertainty (blue line) and with 5\% systematic uncertainty (red line) as a function of predicted probability are shown.}
\label{shap_scoreBP5}
\end{center}
\end{figure}

\begin{table}[!tb]
\begin{center}
\begin{tabular}{|c||c|c|c|c|c|}
\hline
Benchmark & Probability & Signal & Total   & Signal  &  Gain in $\sigma_{ss}$ \\
Points & Score & Yield & Background &  Significance $\sigma_{ss}$ & from  \\
& &  & Yield &  (Sys Unc. = $5\%$) & Cut-based \\
\hline
\hline
\texttt{BP4} & 0.90  & 356.72 & 7.46  & 18.69 (13.52)  & 51\% (34\%)\\
\cline{2-6}
 & 0.96 & 352.86 & 6.25 &  18.62 (13.51) & 51\% (34\%) \\
\hline
\hline
\texttt{BP5} & 0.90  &  71.80 & 0.94  & 8.42 (7.74) & 33\% (29\%)  \\
\cline{2-6}
 & 0.96 & 71.54  & 0.78 & 8.41 (7.74) & 32\% (29\%)  \\
\hline
\hline
\texttt{BP6} & 0.90  & 41.27  & 0.83 & 6.36 (6.05) & 42\% (40\%) \\
\cline{2-6}
 &  0.96 & 41.20 & 0.08 & 6.41 (6.10) & 43\% (41\%) \\
\hline
\hline
\end{tabular}
\caption{ Signal yield, total background yield and the signal significance (without any systematic uncertainity) at the HE-LHC using ML-based algorithm for different probability scores are presented here. The numbers in the parenthesis correspond to $\sigma_{ss}$ with systematic uncertainty $\epsilon = 5\%$. Here the SUSY signal belongs to 
{\tt{Scenario-I.}}}
 \label{tab:ml_27}
\end{center}
\end{table}

\begin{figure}[!htb]
\centering
\input{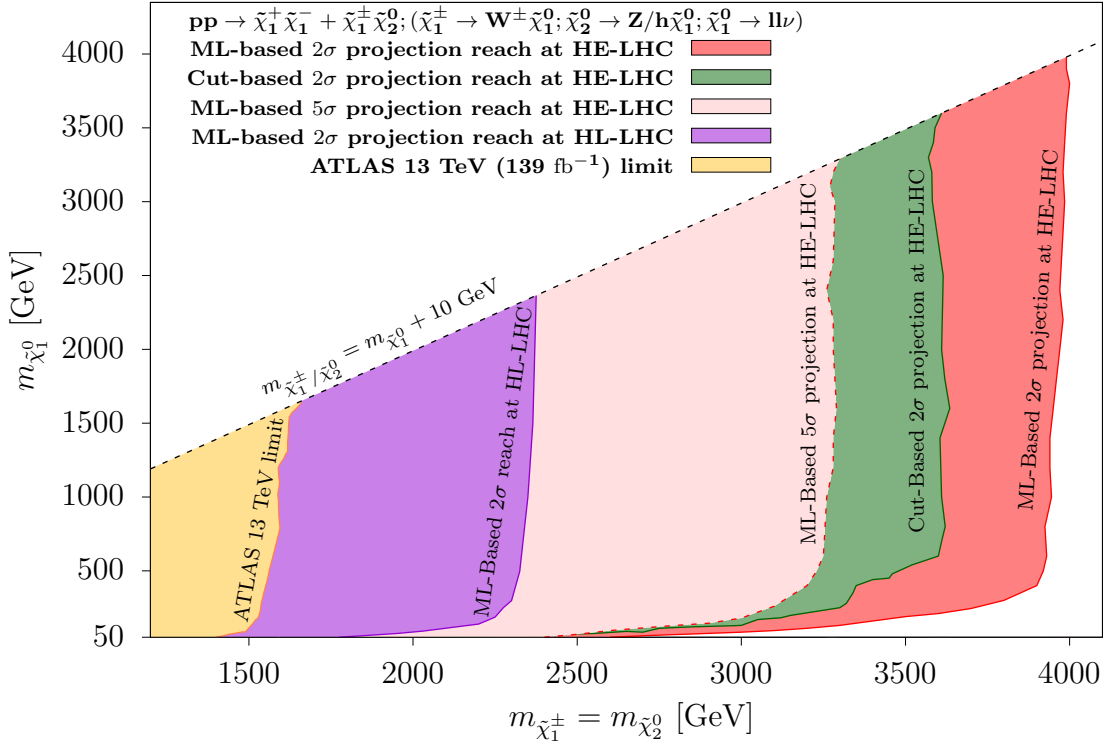}
\caption{ Projected discovery ($5\sigma$) and exclusion ($2\sigma$) regions in the $\mlsptwo-\mlspone$ mass plane at the HE-LHC are presented with light and dark red colors. The green  and dark violet color region represent the  2$\sigma$ reach  obtained by cut-based analysis at the HE-LHC and  ML-based analysis at the HL-LHC respectively. The yellow regions represents the existing limit obtained by the ATLAS collaboration from Run-II data\cite{ATLAS:2021yyr}. }
\label{fig:exclusion_plot_27} 
\end{figure}

We have used a similar ML algorithm as discussed in Sec.~\ref{sec:14_ml} to improve the cut-based analysis results for HE-LHC. For the ML analysis, we have considered the same set of 18 features and followed the same procedure for training, testing, hyper-parameter selection etc. as described in Sec.~\ref{sec:14_ml}. The Shapley feature importance plot for the top 10 kinematic variables and the variation of signal significance as a function of probability score for HE-LHC are presented in Fig.~\ref{shap_scoreBP5}. We observe an almost similar SHAP ranking in features compared to HL-LHC analysis and the most important features are $N_Z$, $m_{eff}$. $p_T^{l_2}$ and $p_T^{l_1}$ for HE-LHC analysis.

In Table.~\ref{tab:ml_27}, the signal yield, total backgrounds yield, signal significance with 0\% and 5\% systematic uncertainty for the representative  benchmark points \texttt{BP4}, \texttt{BP5} and \texttt{BP6} are displayed for probability score 0.90 and 0.96. 
We notice that there are around $\sim$ 30-50\% gains in signal significance compared to the cut-and-count method. The gain is also reflected in the projected exclusion plot (Fig.~\ref{fig:exclusion_plot_27}) displayed  in the LSP-NLSP mass plane. We find that the projected discovery limit reaches upto 3.3 TeV (illustrated in light red color) in ML-based analysis which is around 140 GeV larger than the cut-based reach. Also, the projected exclusion limit extends upto $\sim$ 4.0 TeV (presented by dark red color in Fig.~\ref{fig:exclusion_plot_27}) resulting a 380 GeV enhancement from the traditional cut-and-count method. The dark violet and dark green region correspond to ML-based 
2$\sigma$ reach at the HL-LHC and cut-based  2$\sigma$ reach at the HE-LHC respectively 
which is summarized in Sec~\ref{sec:14_ml} and Sec.~\ref{sec:27_cut}.

\begin{table}[!htb]
\begin{center}
\begin{tabular}{|c|c|c|c|c|} 
    \hline
	Benchmark & \multicolumn{4}{c|}{\textbf{ML} based Signal Significance (Syst. Unc. = 5\%)} \\
	\cline{2-5}
	points & Scenario-I & Scenario-II & Scenario-III & Scenario-IV \\
	\hline
    \texttt{BP4} & 18.69 (13.52) & 12.72 (10.63) & 7.99 (7.36) & 5.21 (4.99)\\
    \hline
    \texttt{BP5} & 8.42 (7.74) & 6.4 (6.09) & 4.21 (4.11) & 2.97 (2.93) \\
    \hline
    \texttt{BP6} & 6.36 (6.05) & 4.98 (4.83) & 3.23 (3.19) & 2.31 (2.29) \\
    \hline\hline
    $m_{\lspone}$  & \multicolumn{4}{c|}{Projected exclusion on $m_{\chonepm}$ at the HE-LHC (Sys. Unc.= 20\%) } \\
	\hline
    1200 & 3940 (3850) & 3790 (3700) & 3450 (3360) & 3200 (3115) \\
    \hline
\end{tabular}
\caption{Comparison of signal significance of benchmark points \texttt{BP4}, \texttt{BP5} and \texttt{BP6} for different model scenarios 
(defined in Table~\ref{tab:br2}) with  0\% (5\%) systematic uncertainty are shown here. The numbers in last row represents the projected 95\% C.L. 2$\sigma$ exclusion limits on NLSP masses for a fixed 1200 GeV LSP with 0\% (20\%) systematic uncertainty. Here all the masses are in GeV.}
\label{tab:scenario_compare_27}
\end{center}
\end{table}

Now we proceed to study the effect of choosing different couplings or SUSY scenarios as defined in Table.~\ref{tab:br2}. The signal significance for \texttt{BP4}, \texttt{BP5} and \texttt{BP6} and the projected 2$\sigma$ exclusion limits on $m_{\tilde{\chi}_1^{\pm}}$ for fixed $m_{\tilde{\chi}_1^0}$ = 1200 GeV are listed in Table.~\ref{tab:scenario_compare_27}. As expected the maximum reach is obtained for \texttt{Scenario-I} (3.94 TeV) and \texttt{Scenario-IV} corresponds to the least sensitive compare to others (3.2 TeV). The effect of systematic uncertainty are also displayed  in Table.~\ref{tab:scenario_compare_27}.  
It may be noted that  the 
projected exclusion limits on $\mchonepm$ as mentioned in Table~\ref{tab:scenario_compare_27} 
improve by $\sim$ 50-60 GeV if the LSP-NLSP mass gap is 
$\sim$ 10 GeV.


\section{Conclusion}
\label{sec:conclusion}
Supersymmetry remains one of the most highly motivated beyond the SM scenario both theoretically and phenomenologically. In the absence of any significant excess over the SM from the experimental results, it is important to study the existing models under existing data  and assess how much of the relevant parameter space can be probed at the highest luminosity of the LHC. In the process, the canonical search techniques are to be put to comparison with the new tools available at our disposal at present to assess how much we can improve on the existing sensitivities. The gaugino sector of the supersymmetry has diverse phenomenological implications and hence is of very high interest. The gaugino sector of the R-parity conserving MSSM has been studied exhaustively in this regard while the various R-parity violating scenarios have not been explored to that extent. In this work, we have chosen a multilepton ($N_l \geq 4$ with $l\equiv e,~\mu$) final state to assess the discovery and exclusion reach of the high luminosity LHC as well as the high energy LHC in terms of the gaugino masses. We have compared the sensitivity of probing the parameter space through traditional cut-based method and machine learning based method. Our results clearly show that one can expect a gain of upto $43\%$ and $51\%$ in signal significance using the gradient boosted decision tree algorithm over that of the cut based analyses in the context of the HL-LHC and HE-LHC respectively. This leads to a far better reach in the exclusion and discovery limits in the wino-bino mass plane. 
For scenarios with nonzero $\lambda_{121}$ and/or $\lambda_{122}$,  the projected discovery reach at the HL-LHC, obtained by us are $\sim$ 1.99 TeV and $\sim$ 1.87 TeV in ML-based and cut-based methods respectively. Similarly, the projected exclusion curve reaches upto $\sim$ 2.37 TeV and $\sim$ 2.18 TeV respectively. At the HE-LHC, ML-based method provides even better sensitivity. Our projected $5\sigma$ discovery sensitivity reaches upto $\sim$ 3.3 TeV in ML-based analysis which is $\sim$ 140 GeV larger than that of the cut-based reach. The projected exclusion limit reaches $\sim$ 4 TeV which is an improvement by $\sim$ 380 GeV over the corresponding cut-based analysis.
Apart from the scenarios with nonzero $\lambda_{121}$ and/or $\lambda_{122}$, 
we also discuss the possibility of three other scenarios, derived from the remaining 
seven nonzero single $\lambda_{ijk}$ couplings, with varied $\tau$ lepton multiplicity in the final state.
With specific choices of benchmark points, we show how the sensitivities vary for these different scenarios for $N_l \geq 4$ final state.


\newpage
\noindent \textbf{Acknowledgments}

\noindent The authors would like to acknowledge Biplob Bhattacherjee and Camellia Bose for fruitful discussions regarding machine learning analysis.





\bibliography{ref_collider}

\newpage

\appendix
\section{Background cross-sections at the HL-LHC}
\label{appendix1}

\begin{table}[!htb]
\begin{center}
\begin{tabular}{|c|c|c|c|c|}
\hline
\multirow{4}{*}{\shortstack{Background}} & \multirow{4}{*}{\shortstack{Cross-section \\ Order [Ref]}} & \multirow{4}{*}{\shortstack{Cross-section \\ $\sigma$ (fb)}} & \multirow{4}{*}{\shortstack{$\sigma^\prime=$ \\ $\sigma\times Br.(4l)$ \\ (fb)}} & \multirow{4}{*}{\shortstack{ Yield at the \\ HL-LHC after \\ generation level \\ cut $p_T^{l_1} >$ 100 GeV }}  \\
& &  &  & \\
& & & & \\
& & & & \\
\hline
$ZZ$ + jets& NNLO \cite{Cascioli:2014yka} & $18.77\times 10^3$ &  86.79 & 53640 \\
\hline
$t\bar{t}Z$ + jets & NLO \cite{Azzi:2019yne} & $1.018\times 10^3$ & 3.35 & 5931 \\
\hline
$WWZ$ + jets & NLO \cite{Binoth:2008kt} & 181.7 & 0.598 & 1059 \\
\hline
$WZZ$ + jets & NLO \cite{Alwall:2014hca} & 64 & 0.296 & 516 \\
\hline
$ZZZ$ + jets & NLO\cite{Binoth:2008kt} & 15.3 & 0.197 & 282 \\
\hline
$h$  & N3LO QCD& $54.72\times 10^3$ & 7.037 & 3042 \\
(via GGF) & +NLO EW\cite{xsec_twiki} & & & \\
\hline
$hjj$ & NLO \cite{Alwall:2014hca} & $6.42\times 10^3$ & 0.826 & 381 \\
\hline
$Wh$ + jets & NNLO QCD& $1.498\times 10^3$ & 0.193 & 78 \\
& +NLO EW \cite{xsec_twiki} & & & \\
\hline
$Zh$ + jets & NNLO QCD& 981 & 0.126 & 50.1 \\
& +NLO EW \cite{xsec_twiki} & & & \\
\hline
\end{tabular}
\caption{Cross-sections for various relevant SM backgrounds at the 14 TeV LHC are shown here. The corresponding number for $4l$ final states by 
multiplying the appropriate branching ratios and the yields at the HL-LHC after applying a 
generation level cut $p_T^{l_1} > 100$ GeV are presented in the last two column respectively.}
\label{tab:cs_bkg_14}
\end{center}
\end{table}

\vspace{-1cm}

\section{Signal cross-sections at the HL-LHC and HE-LHC}
\label{appendix2}
\begin{table}[h]
\begin{center}
\begin{tabular}{|c|c||c|c|}
\hline
\multirow{3}{*}{\shortstack{Signal \\ ($m_{\tilde\chi_1^{\pm}/\tilde\chi_2^{0}},m_{\tilde\chi_1^{0}}$) GeV}}  & \multirow{3}{*}{\shortstack{Cross-section \\ $\sigma$ (fb) at 14 TeV \\ (NLO+NLL)}} & \multirow{3}{*}{\shortstack{Signal \\ ($m_{\tilde\chi_1^{\pm}/\tilde\chi_2^{0}},m_{\tilde\chi_1^{0}}$) GeV}}  & \multirow{3}{*}{\shortstack{Cross-section \\ $\sigma$ (fb) at 27 TeV\\ (NLO+NLL)}}\\
& &   &\\
& &   &\\
\hline
\multirow{2}{*}{\shortstack{BP1  \\ (1600,250) }}  & \multirow{2}{*}{0.107} & \multirow{2}{*}{\shortstack{BP4  \\ (2300,250) }} & \multirow{2}{*}{0.219} \\
 & &  &  \\
\hline
\multirow{2}{*}{\shortstack{BP2  \\ (1800,800)}}  & \multirow{2}{*}{0.042} & \multirow{2}{*}{\shortstack{BP5  \\ (2900,1200)}}  & \multirow{2}{*}{0.042} \\
 &  &  &\\
\hline
\multirow{2}{*}{\shortstack{BP2  \\ (1950,1850)}}  & \multirow{2}{*}{0.019} & \multirow{2}{*}{\shortstack{BP6  \\ (3100,3000)}}  & \multirow{2}{*}{0.025}\\
 & &  &\\
\hline
\end{tabular}
\caption{NLO+NLL cross-sections for the 14 TeV \& 27 TeV LHC for signal benchmark points 
obtained using\cite{Fuks:2013vua}.}
\label{tab:cs_sig}
\end{center}
\end{table}

\newpage 

\section{Background cross-sections at the HE-LHC}
\label{appendix3}

\begin{table}[h]
\begin{center}
\begin{tabular}{|c|c|c|c|c|}
\hline
\multirow{4}{*}{\shortstack{Background}} & \multirow{4}{*}{\shortstack{Cross-section \\ Order [Ref]}} & \multirow{4}{*}{\shortstack{Cross-section \\ $\sigma$ (fb)}} & \multirow{4}{*}{\shortstack{$\sigma^\prime=$ \\ $\sigma\times Br.(4l)$ \\ (fb)}} & \multirow{4}{*}{\shortstack{ Yield at the \\ HE-LHC after \\ generation level \\ cut $p_T^{l_1} >$ 150 GeV }}  \\
& &  &  & \\
& & & & \\
& & & & \\
\hline
\hline
$ZZ$ + jets& NNLO \cite{Azzi:2019yne} & $44.52\times 10^3$ &  205.86 & 53730 \\
\hline
$t\bar{t}Z$ + jets & NLO \cite{Azzi:2019yne} & $4.9\times 10^3$ & 16.126 & 16740 \\
\hline
$WWZ$ + jets & NLO \cite{Alwall:2014hca} & 573.04 & 1.886 & 2355 \\
\hline
$WZZ$ + jets & NLO \cite{Alwall:2014hca} & 197.1 & 0.911 & 1116 \\
\hline
$ZZZ$ + jets & NLO \cite{Alwall:2014hca} & 41.77 & 0.537 & 429 \\
\hline
$h$ & N3LO QCD & $146.65\times 10^3$ & 18.859 & 6675 \\
(via GGF) & +NLO EW  \cite{xsec_twiki} & & & \\
\hline  
$hjj$ & NLO \cite{Alwall:2014hca} & $15.977\times 10^3$ & 2.05 & 308 \\
\hline
$Wh$ + jets & NNLO QCD & $3.397\times10^3$ & 0.437 & 63 \\
& +NLO EW\cite{xsec_twiki} & & & \\
\hline
$Zh$ + jets & NNLO QCD & $2.463\times10^3$ & 0.317 & 45 \\
& +NLO EW \cite{xsec_twiki} & & & \\
\hline
\end{tabular}
\caption{Cross-sections at the 27 TeV LHC for various SM backgrounds considered in 
this analysis are shown here. Conventions are same as in Table.~\ref{tab:cs_bkg_14}. The last column represents the yields at the HE-LHC after applying a generation level cut $p_T^{l_1} > 150$ GeV.}
\label{tab:cs_bkg_27}
\end{center}
\end{table}

\end{document}